\pdfoutput=1
\def\chronosarxiv{}
\documentclass[conference,compsoc]{IEEEtran}

\input{glyphtounicode}
\pdfglyphtounicode{circumflex}{005E}
\usepackage{amsmath}
\usepackage{amsthm}

\usepackage{amssymb}
\usepackage{booktabs}
\usepackage{array}
\usepackage{tabularx}
\usepackage[table]{xcolor}
\usepackage{colortbl}

\usepackage{graphicx}
\usepackage{tikz}
\usetikzlibrary{arrows.meta,positioning,decorations.pathreplacing,calc,fit,backgrounds,patterns.meta}
\usepackage{pgfplots}
\usepackage{pgfplotstable}
\pgfplotsset{compat=1.18}
\usepgfplotslibrary{groupplots}
\definecolor{ChInk}{HTML}{202A32}
\definecolor{ChInkSoft}{HTML}{59636A}
\definecolor{ChGuide}{HTML}{9BA1A6}
\definecolor{ChRule}{HTML}{CDD0D2}
\definecolor{ChPaper}{HTML}{F5F6F7}

\definecolor{ChRuntime}{HTML}{2378B5}
\definecolor{ChRuntimeTint}{HTML}{EDF4F9}
\definecolor{ChRuntimeStrong}{HTML}{D7E7F2}
\definecolor{ChAttack}{HTML}{A3550A}
\definecolor{ChAttackTint}{HTML}{FBE6C7}
\definecolor{ChAttackStrong}{HTML}{F8DAAA}
\definecolor{ChLabel}{HTML}{B8407C}
\definecolor{ChLabelTint}{HTML}{F6E8EF}
\definecolor{ChLabelStrong}{HTML}{F1D9E5}

\definecolor{ChBandUser}{HTML}{E1EDF7}
\definecolor{ChBandUserEdge}{HTML}{9DBFDB}
\definecolor{ChBandKernel}{HTML}{E8EBEE}
\definecolor{ChBandKernelEdge}{HTML}{AAB3BC}
\definecolor{ChBandBelow}{HTML}{D9DEE3}
\definecolor{ChBandHost}{HTML}{F2F0EC}
\definecolor{ChBandChronos}{HTML}{FCEFDC}
\definecolor{ChBandChronosEdge}{HTML}{E2B77C}

\tikzset{
  ch/.style={font=\footnotesize, line width=0.4pt, draw=ChInk, text=ChInk,
    >={Stealth[length=3.4pt, width=2.8pt]}, line cap=round, line join=round},
  ch box/.style={draw=ChRule, fill=white, inner xsep=3pt, inner ysep=2.4pt, align=center},
  ch input/.style={ch box, fill=ChPaper},
  ch runtime/.style={ch box, draw=ChRuntime, fill=ChRuntimeTint},
  ch attack/.style={ch box, draw=ChAttack, fill=ChAttackTint},
  ch label/.style={ch box, draw=ChLabel, fill=ChLabelTint},
  ch note/.style={font=\scriptsize, text=ChInkSoft, align=center, inner sep=1pt},
  ch panel/.style={font=\sffamily\bfseries\footnotesize, text=ChInk, anchor=base west, inner sep=0pt},
  ch lane name/.style={font=\scriptsize, text=ChInkSoft, anchor=base east, inner sep=0pt},
  ch lane/.style={draw=ChGuide, line width=0.45pt, ->},
  ch event/.style={circle, fill=#1, draw=white, line width=0.35pt, inner sep=0pt, minimum size=4.4pt},
  ch event/.default=ChInk,
  ch flow/.style={->, line width=0.8pt},
  ch link/.style={->, line width=0.4pt},
  ch held/.style={line width=2.6pt, draw=ChRuntime, line cap=butt},
  ch break/.style={draw=white, line width=1.6pt, line cap=butt},
  ch step/.style={circle, draw=ChRuntime, fill=ChRuntimeTint, text=ChRuntime,
    font=\sffamily\bfseries\scriptsize, inner sep=0.6pt, minimum size=9pt},
  ch point/.style={circle, fill=#1, text=white, font=\sffamily\bfseries\fontsize{5.6}{6}\selectfont,
    inner sep=0pt, minimum size=7.8pt},
  ch point/.default=ChInk,
  ch phase/.style={rectangle, rounded corners=1.8pt, draw=ChInk, fill=white, text=ChInk, line width=0.4pt,
    font=\sffamily\bfseries\fontsize{5.6}{6}\selectfont, inner sep=0pt, minimum width=8.2pt, minimum height=7.4pt},
  ch case/.style={rectangle, draw=ChInk, fill=white, text=ChInk, line width=0.4pt,
    font=\sffamily\bfseries\fontsize{5.6}{6}\selectfont, inner sep=0pt, minimum size=7.8pt},
  ch band user/.style={fill=ChBandUser, draw=ChBandUserEdge, line width=0.5pt},
  ch band kernel/.style={fill=ChBandKernel, draw=ChBandKernelEdge, line width=0.5pt},
  ch band below/.style={fill=ChBandBelow, draw=ChBandKernelEdge, line width=0.5pt},
  ch band chronos/.style={fill=ChBandChronos, draw=ChBandChronosEdge, line width=0.5pt},
  ch band title/.style={font=\sffamily\bfseries\fontsize{6.5}{7.6}\selectfont, inner sep=0pt},
}

\DeclareRobustCommand{\chpt}[2][ChInk]{%
  \tikz[baseline=(chpt.base)]\node[ch point=#1] (chpt) {#2};}

\DeclareRobustCommand{\chstep}[1]{%
  \tikz[baseline=(chstep.base)]\node[ch phase] (chstep) {#1};}

\DeclareRobustCommand{\chyes}{\tikz[baseline=-0.6ex]\fill[ChInk] (0,0) circle (2.1pt);}
\DeclareRobustCommand{\chpart}{\tikz[baseline=-0.6ex]{%
  \fill[ChInk] (0,0) -- (0,-2.1pt) arc[start angle=-90, end angle=90, radius=2.1pt] -- cycle;
  \draw[ChInk, line width=0.45pt] (0,0) circle (1.875pt);}}
\DeclareRobustCommand{\chno}{\tikz[baseline=-0.6ex]\draw[ChInk, line width=0.45pt] (0,0) circle (1.875pt);}

\pgfplotsset{
  chronos plot/.style={
    scale only axis,
    axis line style={ChGuide, line width=0.4pt},
    every tick/.style={ChGuide, line width=0.4pt},
    major tick length=2pt, minor tick length=1.2pt,
    tick align=outside, tick pos=left,
    tick label style={font=\scriptsize, text=ChInkSoft},
    label style={font=\footnotesize, text=ChInk},
    xmajorgrids, ymajorgrids,
    major grid style={ChRule!65, line width=0.3pt},
    axis x line*=bottom, axis y line*=left,
    clip=false,
    /pgf/number format/assume math mode=true,
    /pgf/number format/1000 sep={,},
  },
  discard if not/.style 2 args={
    filter discard warning=false,
    x filter/.append code={
      \getthisrow{#1}{\chFilterValue}%
      \edef\chFilterValue{\chFilterValue}%
      \edef\chFilterWant{#2}%
      \ifx\chFilterValue\chFilterWant\else\fi
    },
  },
}

\newcommand{\tabgroup}[2]{\multicolumn{#1}{@{}l}{\rule{0pt}{2.4ex}\itshape #2}\\}

\usepackage{cite}
\usepackage{url}
\usepackage[hidelinks,breaklinks=true]{hyperref}
\usepackage{microtype}
\hypersetup{
  pdftitle={Taking Time Out of TOCTOU: From Race Reproduction to Security Discovery},
  pdfauthor={},
  pdfsubject={Deterministic reproduction, regression testing, and commit bisection for security races},
  pdfkeywords={TOCTOU races, container security, deterministic reproduction, security regression testing, commit bisection}
}
\ifdefined\chronosarxiv
\hypersetup{pdfauthor={Hasnain Naeem and Ryan Henry}}
\fi

\usepackage{listings}
\DeclareTextSymbolDefault{\textquotedbl}{T1}
\colorlet{yamlkeycolor}{ChInk}
\colorlet{yamlvalcolor}{ChInkSoft}
\colorlet{yamlcomcolor}{ChGuide}
\newcommand\YAMLcolonstyle{\color{yamlvalcolor}\mdseries}
\newcommand\YAMLkeystyle{\color{yamlkeycolor}\bfseries}
\newcommand\YAMLvaluestyle{\color{ChInk}\mdseries}
\lstdefinelanguage{yaml}{
  keywords={true,false,null},
  keywordstyle=\color{yamlvalcolor}\bfseries,
  sensitive=false,
  comment=[l]{\#},
  commentstyle=\color{yamlcomcolor}\itshape,
  morestring=[b]',
  morestring=[b]",
  stringstyle=\color{yamlvalcolor},
  basicstyle=\footnotesize\ttfamily\YAMLkeystyle,
  moredelim=[l][\color{yamlcomcolor}\itshape]{\#},
  moredelim=**[il][\YAMLcolonstyle{:}\YAMLvaluestyle]{:},
}
\lstdefinestyle{chronosyaml}{
  language=yaml,
  upquote=true,
  showstringspaces=false,
  columns=fullflexible,
  keepspaces=true,
  breaklines=true,
  breakatwhitespace=true,
  frame=leftline,
  framerule=0.6pt,
  rulecolor=\color{ChRule},
  xleftmargin=1.1em,
  aboveskip=0.7em,
  belowskip=0.3em,
}

\newcolumntype{Y}{>{\raggedright\arraybackslash}X}
\makeatletter
\g@addto@macro\UrlBreaks{\do\.\do\_\do\-\do\=\do\:\do\/}
\makeatother
\newcommand{\codebreak}[1]{{\ttfamily\nolinkurl{#1}}}
\newcommand{\renameex}{\texttt{renameat2\allowbreak(RENAME\_\allowbreak EXCHANGE)}}
\providecommand{\Description}[1]{}

\newtheorem{proposition}{Proposition}

\newcommand{\chNote}[2]{\par\noindent\hangindent=1.1em\hangafter=1\makebox[1.1em][l]{$^{#1}$}#2\strut\par}

\begin{document}
\pagestyle{plain}

\title{Taking Time Out of TOCTOU: From Race Reproduction to Security Discovery}

\ifdefined\chronosarxiv
\author{%
  \IEEEauthorblockN{Hasnain Naeem \qquad Ryan Henry}
  \IEEEauthorblockA{University of Calgary, Calgary, Alberta, Canada\\
    \texttt{hasnain.naeem@ucalgary.ca} \qquad
    \texttt{ryan.henry@ucalgary.ca}}
}
\else
\author{}
\fi

\maketitle
\thispagestyle{plain}

\begin{abstract}
  Time-of-check-to-time-of-use (TOCTOU) races are hard to reproduce and lost races can make vulnerable programs
    look fixed.
  We present Chronos, a framework that separates intervention timing from verdict evidence and builds
    reusable one-trial security regression oracles from declared races for deterministic reproduction, commit bisection and CI.
  Race Specifications compose reusable hold engines, actions, observers and outcome rules.
  When the declared hold is reached and the attack completes while held, Chronos orders it before the use
    under blocking and interference assumptions.
  Chronos uses seccomp notifications to enforce declared hold--attack--release ordering at syscall boundaries
    on stock Linux; ptrace and a KVM guest gate provide holds at userspace and kernel instruction sites.
  One decision procedure labels every mechanism's runs: fixed requires positive, case-specific evidence of
    preservation, justified refusal, omission under checked coverage, or a declared rule; missing or contradictory
    required evidence remains indeterminate.
  On known-vulnerable builds, effect-only checkers made over 800 false-fixed classifications in failed-intervention trials; Chronos made none.
  All 17 syscall-held cases and four of six instruction-held cases had vulnerable and fixed labels.
  Two kernel cases remain one-sided.
  All 27 historical bisections recovered the expected boundaries.
  For runc's masked-path bisection with one victim per launch, fixed retry and Git Bayesect took medians of 475 and
    398 natural launches; Chronos used seven held trials including endpoint checks.
  Since a label means the same however a fix is written, we also ran the specifications and attack variants on
    releases, distribution packages and other runtimes. There, Chronos helped us uncover three issues: incomplete fixes
    in chrony and rpm, and a Debian rpm package its tracker listed as fixed.
\end{abstract}


\section{Introduction}
\label{sec:intro}

Security races are notoriously difficult to reproduce. Practitioners and researchers rely on repeated trials,
  race-specific amplification,
  or controlled interleavings to expose them.
To demonstrate CVE-2021-30465, its discoverer raced 19 victim containers against four swap loops,
  after a smaller setup always failed; 6 of the 19 escaped~\cite{champetier2021runc}.
Fix validation, regression testing and bisection need to distinguish a fixed revision from a vulnerable one whose run
  missed the race.
For a bug that reproduces on every run, all three can run automatically, but races caused 12 of 20 sampled
  reproduction failures in syzbot's cause bisection (\S~\ref{sec:bisect}).
Fixes are often incomplete: 17 of the 41 zero-days exploited in the wild in 2022 were variants of previously reported
  vulnerabilities, and at least half of those found in the first half of that year could have been prevented by more
  comprehensive patching and regression tests~\cite{stone2023zerodays,stone2022zerodays}.
In a study of 429 container-runtime bugs, 78.79\% evaded existing tests for lack of drivers, oracles or effective
  cases~\cite{yu2024bugsinpods}.

\begin{figure*}[t]
\centering
\resizebox{\textwidth}{!}{%
\begin{tikzpicture}[ch, x=1cm, y=1cm,
  lbl/.style={font=\scriptsize, align=center, inner sep=0.8pt},
  up/.style={lbl, anchor=south, yshift=1.8pt},
  down/.style={lbl, anchor=north, yshift=-1.8pt},
  outcome/.style={ch label, font=\scriptsize, text=ChLabel, minimum width=1.3cm,
    minimum height=0.74cm, inner xsep=1pt, align=center},
  sig/.style={->, line width=0.45pt},
]
\def\la{2.05}\def\lb{0.95}
\draw[ChRule] (9.05,0.05) -- (9.05,2.95);
\draw[ChRule] (0,-0.02) -- (18.1,-0.02);

\begin{scope}
  \node[ch panel] at (0,2.75) {(a) Natural: the swap must complete inside $W$};
  \node[ch lane name] at (0.95,\la) {runc};
  \node[ch lane name] at (0.95,\lb) {swapper};
  \begin{scope}[shift={(1.55,0)}, x=0.098cm]
    \fill[ChAttackTint] (0,\lb-0.2) rectangle (27.924,\la+0.06);
    \node[lbl, text=ChAttack] at (13.962,1.3) {$W$};
    \draw[ch lane] (-1.1,\la) -- (57.0,\la);
    \draw[ch lane] (-1.1,\lb) -- (57.0,\lb);
    \fill[pattern={Lines[angle=45, distance=2.2pt, line width=0.35pt]}, pattern color=ChAttack]
      (-0.9,\lb-0.075) rectangle (56.5,\lb+0.075);
    \node[down, text=ChAttack, fill=white, inner xsep=2pt] at (41.0,\lb-0.08) {\texttt{renameat2} swap loop};
    \node[ch event=ChRuntime] at (0,\la) {};
    \node[up, text=ChRuntime] at (0,\la) {SecureJoin check\\($t_{\text{check}}$)};
    \draw[ChRuntime, line width=0.7pt] (14.368,\la) -- (53.408,\la);
    \draw[ChRuntime, line width=0.7pt] (14.368,\la-0.07) -- (14.368,\la+0.07)
      (53.408,\la-0.07) -- (53.408,\la+0.07);
    \node[ch event=ChRuntime] at (27.924,\la) {};
    \node[up, text=ChRuntime] at (38.0,\la) {kernel \texttt{path\_mount} entry};
    \node[down, ch note, anchor=north east, inner xsep=0pt] at (14.368,\la) {p5 14.4};
    \node[down, ch note, anchor=north, inner xsep=0pt, xshift=6pt] at (27.924,\la) {median 27.9 (1{,}000 runs)};
    \node[down, ch note] at (53.408,\la) {p95 53.4};
    \foreach \t in {0,10,20,30,40} {
      \draw[ChGuide] (\t,0.40) -- (\t,0.47);
      \node[ch note, anchor=north] at (\t,0.39) {\t};
    }
    \draw[ChGuide] (50,0.40) -- (50,0.47);
    \node[ch note, anchor=north] at (50,0.39) {50\,\textmu s};
  \end{scope}
  \node[outcome] at (8.19,1.5) {exposed in\\603/2{,}000};
\end{scope}

\begin{scope}[shift={(9.35,0)}]
  \node[ch panel] at (0,2.75) {(b) Held: $W\!\to\!\infty$ while \texttt{mount()} waits};
  \node[ch lane name] at (0.95,\la) {runc};
  \node[ch lane name] at (0.95,\lb) {Chronos};
  \fill[ChAttackTint] (2.35,\lb-0.2) rectangle (4.7,\la+0.06);
  \fill[ChAttackTint] (4.95,\lb-0.2) rectangle (6.0,\la+0.06);
  \node[lbl, font=\footnotesize, text=ChAttack] at (3.52,1.42) {$W\!\to\!\infty$};
  \draw[ChGuide, line width=0.45pt] (1.2,\la) -- (4.7,\la) (1.2,\lb) -- (4.7,\lb);
  \draw[ch lane] (4.95,\la) -- (6.75,\la);
  \draw[ch lane] (4.95,\lb) -- (6.75,\lb);
  \draw[ch held] (2.35,\la) -- (4.7,\la);
  \draw[ch held] (4.95,\la) -- (6.0,\la);
  \node[text=ChInkSoft, inner sep=0pt, font=\scriptsize] at (4.825,\la) {$\cdots$};
  \node[text=ChInkSoft, inner sep=0pt, font=\scriptsize] at (4.825,\lb) {$\cdots$};
  \node[ch event=ChRuntime] at (1.35,\la) {};
  \node[up, text=ChRuntime] at (1.35,\la) {check};
  \node[ch event=ChRuntime] at (2.35,\la) {};
  \node[up, text=ChRuntime] at (2.35,\la) {enters \texttt{mount()}\\($t_{\text{entry}}$)};
  \node[ch event=ChRuntime] at (6.0,\la) {};
  \node[ch event=ChRuntime] at (6.3,\la) {};
  \node[up, text=ChRuntime, anchor=south east, xshift=4pt] at (6.3,\la) {resolves the\\path ($t_{\text{resolve}}$)};
  \draw[sig, ChRuntime] (2.35,\la-0.07) -- (2.35,\lb+0.08);
  \node[ch event=ChAttack] at (2.35,\lb) {};
  \node[down, text=ChAttack] at (2.35,\lb) {notification};
  \node[ch event=ChAttack] at (3.9,\lb) {};
  \node[down, text=ChAttack] at (3.9,\lb) {\texttt{renameat2}\\exchange};
  \node[ch event=ChAttack] at (6.0,\lb) {};
  \draw[sig, ChAttack] (6.0,\lb+0.07) -- (6.0,\la-0.08);
  \node[lbl, text=ChAttack, anchor=east] at (5.95,1.5) {continue\\($t_{\text{resume}}$)};
  \node[ch event=ChLabel] at (6.5,\lb) {};
  \draw[sig, ChGuide, dash pattern=on 1.2pt off 1pt] (6.3,\la-0.07) -- (6.5,\lb+0.08);
  \node[down, text=ChLabel] at (6.5,\lb) {$S$ or $F$};
  \node[outcome] at (7.85,1.95) {rc94 redirected\\in 100/100};
  \node[outcome] at (7.85,1.0) {rc95 preserved\\in 50/50};
\end{scope}

\begin{scope}[shift={(0,-0.60)}]
  \node[ch input, anchor=west, minimum height=0.6cm, text width=3.0cm, font=\scriptsize] (sp) at (0.15,0)
    {Race Specification:\\$U$ and $\Delta_{\text{state}}$};
  \node[ch attack, anchor=west, minimum height=0.6cm, text width=2.9cm, font=\scriptsize] (hd) at (4.35,0)
    {hold $U$, apply\\$\Delta_{\text{state}}$, release};
  \node[ch label, fill=white, anchor=west, minimum height=0.6cm, text width=3.1cm, font=\scriptsize] (lb) at (8.45,0)
    {\textcolor{ChLabel}{\textbf{vulnerable}} $\cdot$ \textcolor{ChLabel}{\textbf{fixed}} $\cdot$\\
     \textcolor{ChLabel}{\textbf{indeterminate}}};
  \node[lbl, anchor=west, align=left] (use) at (12.85,0)
    {a one-trial oracle labels a revision:\\a bisection midpoint, or a CI check};
  \draw[->, ChAttack, line width=0.6pt] (sp) -- (hd);
  \draw[->, ChLabel, line width=0.6pt] (hd) -- (lb);
  \draw[->, ChLabel, line width=0.6pt] (lb) -- (use);
\end{scope}
\end{tikzpicture}%
}
\caption{\textbf{Chronos holds open the fleeting race window.}
  It releases the held use only after the declared change completes.
  CVE-2021-30465 in runc; $S$ and $F$ are the evidence behind a vulnerable and a fixed label, and the other symbols are
  as in \S~\ref{sec:formal}.
  The interval in (a) is an upper bound on $W$ (\S~\ref{sec:eval-window}).}
\Description{Two timeline panels for CVE-2021-30465, each with a runc lane above a swapper or Chronos lane and
  outcome boxes at the right, above a strip that shows the one-trial oracle.
  Panel a is drawn to scale in microseconds: after runc's SecureJoin check at t-check, the kernel's path_mount entry
  spreads from 14.4 to 53.4 microseconds with a median of 27.9 over the 1,000 runs whose final check was on the
  destination path itself; a shaded window W shows where a renameat2 swap loop must complete, and host-root content
  was exposed in 603 of 2,000 attempts.
  Panel b shows the same window stretched across a long, broken interval labeled W to infinity while the thread is
  held inside mount() from t-entry to t-resume; Chronos receives the notification, performs one exchange and
  continues the syscall, and the kernel resolves the path at t-resolve. The vulnerable release rc94 mounts the
  canary, redirected in 100 of 100 runs; rc95, which carries the fix 0ca91f44, held and attacked the same way, mounts the checked
  directory through its descriptor, preserved in 50 of 50 runs.
  The strip shows a Race Specification (use and change) feeding a held run that ends vulnerable, fixed or
  indeterminate; one such trial is an oracle that labels a revision for a bisection midpoint or a CI check.}
\label{fig:race-window}
\end{figure*}
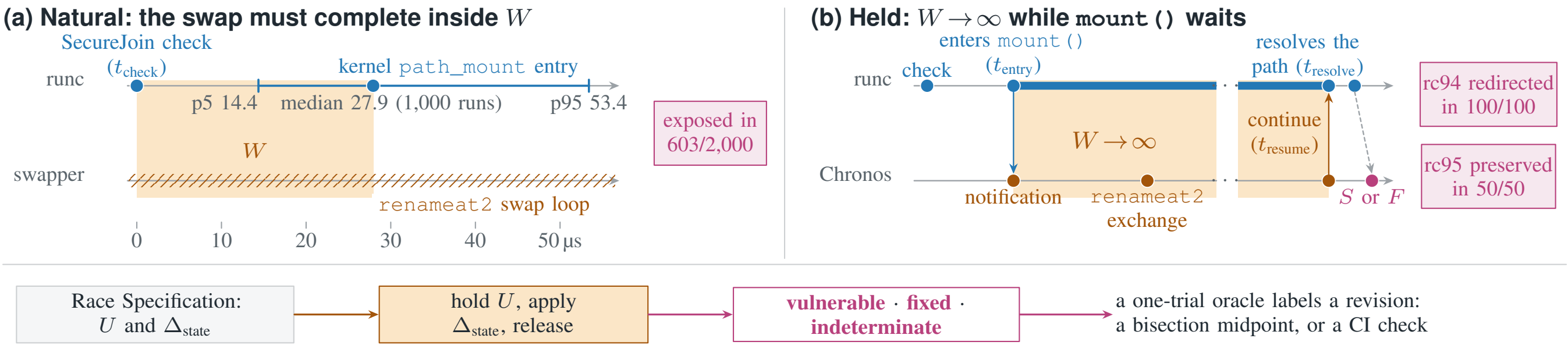%
We evaluate Chronos mainly on time-of-check-to-time-of-use (TOCTOU) races~\cite{bishop1996races} in container
  runtimes,
  where in three of four cases a userspace path check precedes a pathname \texttt{mount(2)} use. The kernel does not
  know which object userspace validated, so an attacker can exchange the name with \renameex{} or a bind mount before
  the use and expose host files inside a container.
Our main cases are
  (A)~a mount-destination TOCTOU, CVE-2021-30465;
  (B)~a mount-source TOCTOU, CVE-2025-23359;
  (C)~the \texttt{/proc} symlink regression, CVE-2023-27561;
  and (D)~the masked-path bug, CVE-2025-31133.

The natural window $W$ ends when the kernel resolves the use path; its median in CVE-2021-30465 is at most
  $28\,\mu$s (Fig.~\ref{fig:race-window}).
Success depends on scheduling and attack setup: measured baseline rates for an \texttt{access()}/\texttt{open()} race
  ranged from about $6.7\times10^{-7}$ to 0.12 across four systems~\cite{deanhu04}, while filesystem mazes and syscall
  synchronization raised another attack above 90\%~\cite{borisov2005}.
Such engineering gives no reusable one-trial regression oracle.
A test need not race on the attacker's terms.
A fixed delay chosen for one change on one host can expire before the change finishes (\S~\ref{sec:eval-window}).
The test controller can instead hold the use, which the attacker in our threat model cannot do to the victim.
The change itself keeps the attacker's permissions.

Chronos targets the remaining gap to 100\% reproduction for the declared race:
  it freezes time and removes the race against it.
  It intercepts the use syscall $U$ just before the kernel processes it and holds still the target thread,
  using \emph{seccomp user-notification} (\texttt{USER\_NOTIF}).
For the mutation, the fleeting race window becomes effectively unbounded ($W\!\to\!\infty$):
  $U$ is released only after the declared mutation (e.g., a swap) completes.

For CVE-2021-30465, the attack succeeded in all 100 runs of the vulnerable version with Chronos and in 603 of
  2{,}000 runs without it (\S~\ref{sec:eval-window}).
A Race Specification declares the use, an attacker change and outcome evidence, which makes a security contract
  testable across revisions: holding the one declared use lets one run decide or end indeterminate, and the
  specification means the same on every revision whose use still matches its patterns.
A separate attacker process makes the change, and Chronos observes the released call's effect, including
  descriptor-based uses in fixed revisions (\S\S~\ref{sec:formal},~\ref{sec:freeze-policy}).
This yields a single-trial oracle under the contract assumptions:
  the use runs only after the declared change completes, whatever the scheduler does;
  the label comes from the selected use's effect or a declared outcome rule, so a fix that changes the mechanism
  but keeps the contract is still labeled fixed;
  and a fixed label needs evidence for the declared repair condition under the case's premises, with missing or
  contradictory required evidence yielding indeterminate.
The closest tools target other races or ask other questions: STING~\cite{sting2012} finds name-resolution flaws through a modified
  kernel; CHESS~\cite{chess2008} and CONCURRIT~\cite{concurrit2013} perform instrumented thread-schedule testing,
  including multi-process system tests in CONCURRIT; and kernel race finders work inside the kernel.

Bisection needs reliable good or bad labels for tested commits.
One wrong verdict sends the search down the wrong half,
  and returns with the wrong commit,
  mis-attributing where a vulnerability was introduced or fixed.
Repeated trials cost more and still yield a statistical verdict, and their number grows without limit as the
  race's rate falls (Proposition~\ref{prop:bisect-bound}).
Meerkat missed an introduction commit several times when a syzbot bug reproduced intermittently, and identifies
  cross-commit reproduction as an open problem~\cite{meerkat2026}.
With one victim per launch, natural bisection's cost followed the race's rate: fixed retry and Git Bayesect (\S~\ref{sec:bisect}) found
  \texttt{runc}'s CVE-2021-30465 fix at medians of 36 and 165 launches and its CVE-2025-31133 fix at 475
  and 398, where Chronos used six and seven held trials.
With the discoverer's 19 victims and four swappers, an amplification specific to that attack, the walks needed
  medians of 8 and 12 launches against six (\S~\ref{sec:eval-bisect}).

Because one held trial labels a build, we ran the specifications, and variants of their attacks, against their
  programs' releases and distribution packages and against \texttt{crun}'s and \texttt{youki}'s releases (RQ4).
Comparing each label with the advisory's or the tracker's record, we found three issues.

The evaluation asks whether one held trial is decisive (RQ1), whether it carries over to other hold engines and bug
  classes (RQ2), whether one held trial per commit suffices for bisection and CI (RQ3), and how race fixes fail (RQ4).

\subsection{Contributions}

Chronos makes a security race reproducible as a single-trial regression oracle whose verdicts rest on evidence:

\begin{enumerate}
\item \textbf{A single-trial oracle with evidence on both sides.}
When the declared hold is reached, the use acts after a completed change (\S~\ref{sec:timing-complete}); its observed
  object, a call never made, a declared rule or the fix's own refusal supplies the label, which keeps its meaning
  across revisions and
  fix forms. Without that evidence the run is indeterminate, so a lost race is never taken for a fix.
Two checkers that read only the effect labeled fixed 800 trials of case A's vulnerable version in which the change
  had not acted before the use, and Chronos labeled none fixed (\S~\ref{sec:eval-window}).

\item \textbf{Race-as-Code.}
One framework composes reusable hold engines, attacker actions, observers, targets and rules around a short Race Specification
  (\S~\ref{sec:offense}), and one decision procedure labels every engine's records, with seccomp
  user-notification the first hold engine.
Of the twenty-three named cases, seventeen use syscall holds, two a userspace instruction hold and four kernel
  instruction holds: twenty-one have both their vulnerable and fixed versions labeled, and two kernel cases remain
  one-sided
  (\S~\ref{sec:eval-further}).

\item \textbf{Bisection and CI.}
\texttt{chronos bisect} checks both ends of the range and runs one trial per tested commit, skipping indeterminate
  midpoints.
All 27 bisections of Table~\ref{tab:bisect-summary} found the expected boundary.
On 30 standard GitHub-hosted CI runners, two specifications labeled all 120 runs correctly
  (\S~\ref{sec:eval-hosted-ci}).

\item \textbf{Incomplete fixes and release sweeps.}
At paths that fixes still resolve by name, a variant of the attack defeated the fixes in chrony and rpm
  (\S~\ref{sec:eval-revealed}).
Debian~12's rpm package, which its tracker listed as fixed, kept rpm's incomplete fix.
Across 194 releases of \texttt{runc}, \texttt{crun} and \texttt{youki}, repetitions reproduced every label, and
  runc's 371 release--specification labels matched advisory status.
With their paths changed and case C's rule naming each runtime's refusal, the same specifications labeled
  \texttt{crun}'s and \texttt{youki}'s releases, and
  bisection named each runtime's repair commits.
They also caught \texttt{runc}'s known reintroduction of an older CVE.
\end{enumerate}

\noindent\textbf{Availability.}
The open-source Chronos framework will be available at \url{https://github.com/hasnainnaeem/chronos}.
An earlier version is at \url{https://github.com/hasnainnaeem/chronos-semester-project}.

\section{Background}
\label{sec:background}

Our main cases come from container runtimes; the other cases of RQ2 (\S~\ref{sec:eval-further}) follow the same
  check--use pattern (\S~\ref{sec:formal-contract}) in other programs and in the kernel.

\subsection{TOCTOU in Container Runtimes}

The runtimes and the NVIDIA hook in our cases use Open Container Initiative (OCI) bundles~\cite{opencontainersSpec} and assemble portions of the container filesystem through privileged pathname operations.
The CVE-2021-30465 exploit~\cite{runcCVE202130465,champetier2021runc} atomically exchanges a mount destination directory with a host symlink via \renameex{} between runc's \texttt{SecureJoin} check and its \texttt{mount()} call.
In CVE-2025-23359~\cite{nvidiaCVE202523359}, the attacker swaps an NVIDIA compatibility library, the source of a bind mount, after the hook validates it and before \texttt{mount()}.
In CVE-2023-27561~\cite{runcCVE202327561}, a regression introduced by the CVE-2021-30465 fix, runc's check that \texttt{/proc} is an ordinary directory examines the resolved path.
A \texttt{/proc} symlink placed in advance therefore passes the check, and runc mounts procfs wherever the
  symlink points.
CVE-2025-31133~\cite{runcCVE202531133} concerns masked paths: runc masks sensitive procfs files such as \texttt{/proc/kcore}, which exposes kernel memory, by mounting \texttt{/dev/null} over them; if an attacker hides \texttt{/dev/null} just before that mount, runc silently leaves the file unmasked.

\subsection{\texttt{openat2} vs.\ \texttt{mount()}}
\label{sec:vfs-gap}

\texttt{openat2()} flags such as \allowbreak\texttt{RESOLVE\_BENEATH} constrain path resolution beneath a
  directory~\cite{LinuxOpenat2}; \texttt{crun} opens mount destinations with \texttt{RESOLVE\_IN\_ROOT} and mounts
  onto the descriptor's \texttt{/proc/self/fd} path.
\texttt{mount(2)} names source and target by pathname, with no such flags, so a runtime that validates a pathname
  and later mounts it has the kernel resolve that name again at use time~\cite{LinuxMount2}.
Even \codebreak{/proc/self/fd/N} is a pathname, though it may refer to an object pinned by a descriptor, and
  \texttt{filepath-securejoin}'s string-returning API cannot stop an attacker from changing path components before
  the caller uses the path.
The fd-based mount API's \codebreak{open_tree()} and \codebreak{move_mount()} take descriptors
  instead~\cite{LinuxOpenTree2,LinuxMoveMount2}.

\section{Threat Model}
\label{sec:threat}

In container setup, our targets are \emph{userspace container runtimes} (\texttt{runc}, \texttt{crun}, \texttt{youki}), the
  NVIDIA container hook (\texttt{nvidia-container-cli}) and kubelet (Appendix~\ref{sec:eval-cases}), all on standard
  Linux kernels.

\subsection{Attacker Capabilities in Container Setup (A1--A4)}
\label{sec:threat-a1a4}

\begin{description}
  \item[A1] The attacker runs arbitrary code inside a container, or supplies a malicious container image.
  \item[A2] Within its UID and mount namespace, the attacker controls the container's filesystem.
  \item[A3] On paths the runtime will later mount, the attacker can create symlinks and perform \renameex{} swaps.
  \item[A4] Using \texttt{/proc} or a predictable setup phase, the attacker can estimate when the runtime reaches setup
    or mount, or it can run a tight swap loop throughout that phase, with no cycle-accurate trace or kernel exploit.
\end{description}

\label{sec:threat-a4}%
A4 matters only for unheld attacks, whose success depends on whether the change completes within the natural
  window.

\subsection{Realistic-Capability Criterion}
\label{sec:realistic-cap}

While Chronos holds the use, the attacker's change must be one that an attacker within the threat model could make
  without Chronos.
The controller uses its host privileges to launch, hold, observe and release the target, never to make the change.
In container setup, every attacker controls only a container, its image or its volumes, and Chronos runs the helper
  that makes the change as that attacker, always without capabilities.

For cases A and B and \texttt{runc}'s CVE-2025-52565, the helper runs as an unprivileged user with write access to
  the shared directory (for B, a shared compat directory; for CVE-2025-52565, the container's \texttt{/dev}, shared by
  a malicious configuration, as the advisory assumes), modeling a peer container that mounts it as a volume.
For case D, the helper runs as uid~0 inside a second container
  that shares only the directory holding \texttt{/dev/null}; case C needs no helper, because its symlink is in the
  image before the check (Table~\ref{tab:oracle-evidence}).

Before releasing the held call, Chronos verifies that the helper ran in this context, and any mismatch makes the
  run indeterminate.
It reads the blocked helper's credentials, capabilities, namespaces and, for a helper entering the target root, that
  root through \texttt{/proc} before letting the helper act.

Under this criterion, a vulnerable label shows that the version has the flaw in the configuration the setup creates.
For a timed race, the hold constructs the after-check, before-use order the attacker can attempt under A3--A4.

Beyond container setup (\S~\ref{sec:eval-further}), each race keeps its report's attacker: the account that owns a
  directory the privileged program writes in (shadow, rpm and Linux-PAM), a local user who can write inside a
  root daemon's module (rsync), a local user who crashes a setuid program and reuses its process ID (apport), a local
  user allowed to run \texttt{sudoedit} (sudo), a client sending requests (Apache), a root program inside the guest
  (QEMU), or a local user running a reproducer (the kernel races; for iommufd, one who can open \texttt{/dev/iommu}).
For chrony, the attacker can already execute code as the service account, and \texttt{chronyc} runs as root.
For Cyrus IMAP, the attacker is the service account that owns the socket directory.
For Docker's copy race, the attacker runs as root inside a container without capabilities.

\subsection{Explicit Non-Goals}
\label{sec:threat-nongoals}

We exclude from the threat model
  attacks on a userspace target through kernel exploits;
  containers that have \texttt{CAP\_SYS\_ADMIN} in the initial user namespace, since such a container can already
  mount host paths directly;
  gaps in the trusted computing base (TCB), such as write access to the OCI bundle's parent directory;
  and pointer-in-syscall TOCTOU vulnerabilities, in which another thread changes a syscall's arguments in memory.

\section{Check--Use Contract Model}
\label{sec:formal}

\begin{figure*}[t]
\centering
\begin{tikzpicture}[ch, x=1cm, y=1cm,
  lbl/.style={font=\scriptsize, align=center, inner sep=0.8pt},
  up/.style={lbl, anchor=south, yshift=1.8pt},
  down/.style={lbl, anchor=north, yshift=-3.2pt},
  guide/.style={draw=ChRule, line width=0.4pt},
  path/.style={draw=ChRuntime, line width=0.9pt, line join=round},
  bar/.style={draw=ChRule, fill=ChPaper, line width=0.4pt},
]
\def\ya{4.70}\def\yr{3.90}\def\ys{3.10}\def\yk{2.30}
\def\tres{13.4}

\foreach \y/\name in {\ya/attacker, \yr/runtime, \ys/syscall entry, \yk/kernel} {
  \draw[guide] (1.95,\y) -- (17.9,\y);
  \node[ch lane name] at (1.8,\y) {\name};
}

\fill[ChAttackTint] (2.6,\ya-0.2) rectangle (\tres,\ya+0.2);
\node[lbl, text=ChAttack] at (8.6,\ya) {window $W$: the change must complete here};
\draw[ChInk, dash pattern=on 2pt off 1.6pt, line width=0.5pt] (\tres,5.12) -- (\tres,1.62);
\node[lbl, anchor=south] at (\tres,5.12) {kernel resolves the path};
\node[ch note, anchor=west, fill=white] at (13.75,\ya) {after resolution: observe or undo};

\draw[path, ->] (1.95,\yr) -- (6.0,\yr) -- (6.0,\ys) -- (11.4,\ys) -- (11.4,\yk)
  -- (16.4,\yk) -- (16.4,\yr) -- (17.9,\yr);
\node[ch event=ChRuntime] at (2.6,\yr) {};
\node[up, text=ChRuntime] at (2.6,\yr) {check $C$};
\node[ch event=ChRuntime] at (6.0,\yr) {};
\node[up, text=ChRuntime] at (6.0,\yr) {calls $U$};
\node[ch event=ChRuntime] at (\tres,\yk) {};
\node[ch event=ChRuntime] at (15.75,\yk) {};
\node[down, text=ChRuntime] at (15.75,\yk) {commits};
\node[ch event=ChRuntime] at (16.4,\yr) {};
\node[up, text=ChRuntime] at (16.4,\yr) {returns};

\node[ch point] at (4.2,\ya) {1};
\node[down] at (4.2,\ya) {calibrated delay\\[-1pt]\textcolor{ChInkSoft}{CARDSHARK~\cite{cardshark2024}}};
\node[ch point] at (4.3,\yr) {2};
\node[down] at (4.3,\yr) {instrumentation\\[-1pt]\textcolor{ChInkSoft}{SysPro~\cite{SysPro2026}}};
\node[ch point=ChAttack] at (6.75,\ys) {3};
\node[down] at (6.75,\ys) {ptrace stop\\[-1pt]\textcolor{ChInkSoft}{measured}};
\draw[ch held] (8.7,\ys) -- (9.55,\ys);
\draw[ChGuide, line width=0.45pt] (9.52,\ys-0.12) -- (9.62,\ys+0.12) (9.68,\ys-0.12) -- (9.78,\ys+0.12);
\node[lbl, text=ChAttack] at (9.15,\ys+0.32) {$W\to\infty$};
\node[ch point=ChAttack] at (8.55,\ys) {4};
\node[down] at (8.55,\ys) {hold at entry\\[-1pt]\textcolor{ChInkSoft}{Chronos}};
\node[ch point=ChAttack] at (10.45,\ys) {5};
\node[down] at (10.45,\ys) {BPF stop\\[-1pt]\textcolor{ChInkSoft}{measured}};
\draw[->, ChAttack, line width=0.55pt, dash pattern=on 0.6pt off 1.2pt]
  (10.52,\ys+0.1) .. controls (12.0,3.62) and (15.2,3.62) .. (16.32,\yr-0.08);
\node[lbl, text=ChAttack, fill=white, inner xsep=1.5pt] at (14.2,3.55) {stop takes effect after commit};
\node[ch point] at (12.0,\yk) {6};
\node[down, xshift=-4.5mm] at (12.0,\yk) {fault or IRQ stall\\[-1pt]\textcolor{ChInkSoft}{DirtyCred~\cite{lin2022dirtycred}, ExpRace~\cite{exprace2021}}};
\end{tikzpicture}
\caption{\textbf{At the syscall level, a pathname race can be held open only before the kernel resolves the path.}
  Numbered points show where tools intervene; ochre points are stops measured here.
  Chronos's syscall hold uses point 4.}
\label{fig:intercept-map}
\end{figure*}
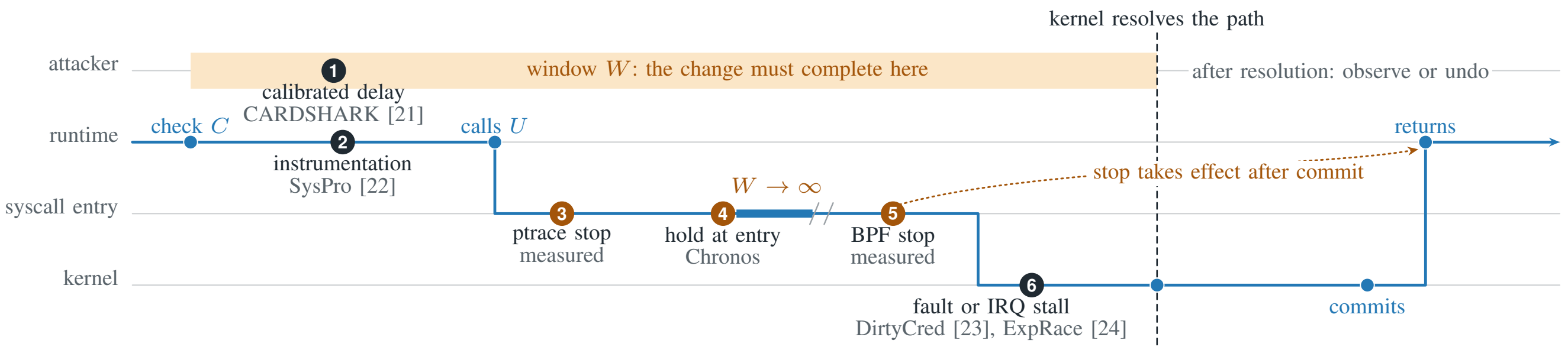

\subsection{Contract}
\label{sec:formal-contract}

For a check--use race, the contract names a use $U$ (such as the syscall \texttt{mount}) on a path $p$,
  an earlier check $C$ of that path (such as \texttt{statx} or a SecureJoin resolution),
  and the attacker's filesystem change $\Delta_{\text{state}}$, made after the check and before the kernel resolves $p$
  for $U$.
The label comes from the object $U$ acted on.
A specification adds an outcome rule only where that object does not decide the label.

The contract's invariant is that $U$ uses only the authority that the preceding validation permitted and performs
  any security effect that the contract requires.
To keep the invariant, a runtime can preserve the validated object with a descriptor,
  resolve the path again under the same policy, or reject the request.
The contract is stated at the use, so a fix that lets $U$ act on the changed object and only afterwards detects the
  change and aborts is labeled vulnerable: a later check cannot undo a use the contract forbids.
Labeling such a fix fixed would need a different contract, stated at the exposure and applied to both labels.
In cases A and B, the use reaches an object the check never validated, and that object decides the label.
Case D has no earlier check, only a required mask that runc omits, and case C needs no change, because the image's
  \texttt{/proc} symlink defeats the check; their specifications state the violation as a rule: the mask is missing,
  or procfs is mounted at the image's path (Table~\ref{tab:oracle-evidence}).

The same contract applies to the non-container races of \S~\ref{sec:eval-further}; for the memory races among
  them, the use is a memory copy or a kernel read, and the change a write or a free.

\subsection{The Hold}
\label{sec:formal-hold}

Let $t_{\text{check}}$ be the end of the check, $t_{\text{entry}}$ the thread's entry into $U$,
  and $t_{\text{resolve}}$ the moment the kernel resolves $p$.
The natural race window is $W = t_{\text{resolve}} - t_{\text{check}}$.
Without help, an attacker must complete $\Delta_{\text{state}}$ inside this window.

Chronos instead enforces the order of these steps.
Through a seccomp user-notification filter, the kernel stops the calling thread at $t_{\text{entry}}$, before running
  $U$~\cite{LinuxSeccompUnotify} (\chpt[ChAttack]{4} in Fig.~\ref{fig:intercept-map}).
The controller releases the thread at $t_{\text{resume}}$, only after $\Delta_{\text{state}}$ has completed:
\[
\mathit{Check} \rightarrow \mathit{Entry/Hold} \rightarrow \Delta_{\text{state}} \rightarrow \mathit{Release}
  \rightarrow \mathit{Resolve}.
\]
Only the calling thread stops, and the hold adds the interval from $t_{\text{entry}}$ to $t_{\text{resume}}$ to the
  window.
We write $W\!\to\!\infty$ because the call to $U$ stays held until the change completes, subject to the controller's
  deadline.
If the change misses its deadline, Chronos lets the call continue without observing its effect, undoes the change,
  and labels the trial indeterminate.
The order needs no particular mechanism, only a \emph{hold point} that suspends the thread before the use reads the
  contested state.
Seccomp's notification gives one at the use syscall's entry, and every main case uses it;
  a specification may instead hold an earlier call of the same thread (\texttt{crun}'s open search, \S~\ref{sec:eval-bisect}),
  and \S~\ref{sec:eval-further} holds a userspace and a kernel instruction.

\subsection{What a Label Means}
\label{sec:freeze-soundness}

At the hold, and again after the change just before release, Chronos records the object (device, inode and type)
  each path argument of the held call names, following a descriptor to its object; after $U$ returns, it observes
  which of the
  two objects $U$ acted on (the \emph{identity check}; Fig.~\ref{fig:chronos-framework-pair-evidence}(b)).

Chronos labels a run \emph{vulnerable} when $U$ acted on the object the change put there (\emph{redirected}),
  or when a rule the specification declares shows the violation (cases C and D).
It labels the run \emph{fixed} when $U$ acted on the object named at the hold even though the change completed
  (\emph{preserved}), so a held use's fixed label rests on what $U$ did under attack.
When the held call comes before the use, preservation alone is not evidence of a fix, because its object may already
  have been pinned before the held checkpoint.

Where the specification declares it, a fixed label can also rest on a rule about the final state after the held
  call was attacked, on the target's refusal with the fix's own message before the use (case C), or on a call never
  made: the run reached the code that would make the use and finished without it (case B).

A refusal rule matches a declared diagnostic when the run ends abnormally, without timeout and without the held
  use; attributing that diagnostic to the relevant pre-use check is a case premise.
Omission records normal completion and no matched use within checked coverage; equivalent forbidden uses must
  fall within that coverage (\S~\ref{sec:limitations}).
After an earlier hold point, as in \texttt{crun}'s open search and the fixed kernel of \S~\ref{sec:eval-further},
  each also needs the change to have completed while the thread was held.
A rule about the final state shows that state when it is read, so it excludes an earlier forbidden effect only if that
  effect would still be visible then; case B's specification states this premise for the compat bind.
A run that refuses with any other message, such as another fix's, is indeterminate, so only refusal evidence needs
  to know the fix.

Every other run is \emph{indeterminate}, for instance when the hold may have missed the call, when setup, the change
  or the undo fails, when the held arguments change, when something other than the attacker changes what they name,
  or when the evidence is contradictory.
We write $S$ for the evidence behind a vulnerable label and $F$ for the evidence behind a fixed one,
  and we call vulnerable and fixed the \emph{definitive} labels.
A label covers only the declared contract: the declared use, its path pattern, and the configuration that the setup
  script creates (\S~\ref{sec:limitations}).
Table~\ref{tab:evidence-contracts} (Appendix~\ref{sec:appendix-evidence-contracts}) distinguishes each recorded
  predicate from the case premises needed to interpret it.

\noindent\textbf{What the observer measures.}
Chronos resolves the held names in the target's own view, and records the object the target's check validated only
  when the specification declares the check call, such as a \texttt{stat}; otherwise the label assumes the held
  object is the one the check validated, a correspondence Proposition~\ref{prop:placement} does not show.
How Chronos measures the object $U$ acted on depends on the syscall family.
For \texttt{mount} (\texttt{mount}, \texttt{open\_tree}, \texttt{move\_mount}), it compares the target's mount table
  after release with a baseline taken before it and asks which held object a newly arrived mount covers; a
  source-focused contract instead compares the object that the arrived mount exposes.
For \texttt{open} (\texttt{open}, \texttt{creat}, \texttt{openat}, \texttt{openat2}), it checks which of the two
  candidate objects the held thread opened.

Three checks tie the measured effect to the held call: Chronos accepts evidence of the call's return only for the
  held notification; the target's other intercepted calls wait until the call's effect is observed (the \emph{effect
  window}, Appendix~\ref{sec:appendix-supervisor}); and names the call looks up the same way must agree, at the hold
  and after the change.
If the observations identify both candidate objects, neither, or none before the deadline, the run is
  indeterminate.
Appendix~\ref{sec:appendix-oracle-ablation} removes five checks of the decision procedure, one at a time.
The case-A row of Table~\ref{tab:oracle-evidence} pairs the redirected vulnerable object with the fixed
  version's preserved object.

\subsection{Single-Trial Guarantee}
\label{sec:timing-complete}

Suppose that the thread calling $U$ stays blocked until release;
  that, until $U$ has consumed its arguments and its effect is observed, no other target thread, signal handler,
  or process besides the declared change modifies the syscall arguments, the path's authority,
  or the state the label depends on;
  and that the run reaches the matching hold and $\Delta_{\text{state}}$ completes before release.
The kernel resolves $p$ only after release, so the change completes before $U$ reads the contested state: in this
  trial the attacker cannot lose the race, and a lost race cannot pass for a fix.
The label then rests on that one trial's evidence: $S$ makes the run vulnerable, $F$ fixed, and anything else
  indeterminate.

The guarantee covers the trial's own path to $U$; a fixed answer about a revision also needs every execution the setup
  admits to reach $U$ the same way, since a program that chooses between a pinned descriptor and a path before the
  check can be safe in one run and vulnerable in the next.
In our cases the setup fixes the configuration and the inputs, so the target takes one path to $U$, which the traces
  below check.

At an instruction hold, a \texttt{ptrace} breakpoint or the kernel gate keeps the thread waiting until release,
  the gate spinning where the task must not sleep.
Only the target's own threads could rewrite the held arguments: no attacker can, since
  pointer-in-syscall races are a non-goal (\S~\ref{sec:threat-nongoals}), and Chronos rereads the arguments just
  before release.
A rewrite after that reread cannot be observed in a multithreaded target, such as runc, so
  Chronos marks such captures racy.
Go gives each of runc's held paths its own buffer, which no other thread references; across 27 groups of
  repeated racy runs of one binary (612 runs), no group gave two labels.

\begin{proposition}[Placement]
\label{prop:placement}
Fix the build, the initial state, the inputs and the target's other choices, and take the premises above.
Suppose that moving the whole change earlier, past each step between the check and $U$, keeps the run legal and
  reaches the
  same complete state, including any time, memory or file state the target reads later.
Then $U$ acts on the same objects wherever inside the window the change completes.
A fixed label from preservation is then correct for every placement of the change inside the window.
\end{proposition}
Whether a natural run can complete the change inside the window is a separate question; the proposition concerns only
  where in the window the change completes (proof in Appendix~\ref{sec:appendix-placement}).
The guarantee and the proposition take $S$ and $F$ as given.

On every run, Chronos checks the assumptions it can observe (the matching hold, the completed change, and the held
  arguments and names just before release), and a failed check makes the run indeterminate (\S~\ref{sec:lifecycle}).
Traces check the system-call part of Proposition~\ref{prop:placement}'s premise at the endpoints and at the midpoints
  that hold a call and make the change, 36 of the 47 in the eight runc and NVIDIA searches.
In none of them does a target call between the check and $U$ touch the changed name, its object or its directory.
In fixes that pin a descriptor, the check is on that descriptor (Table~\ref{tab:oracle-evidence}), unless it is
  opened by name and never checked, as in \texttt{crun}'s \texttt{afeb0654} (\S~\ref{sec:eval-bisect}); case
  D's fix adds a checked one.

A fixed label from a refusal or a call never made comes from a run that never makes the use.
When the run reaches no hold point either, Chronos makes no change, and the label holds for every placement of the
  change when the refusal or the missing call does not depend on the change: case C makes no change, and case B's
  fixed hook skips the compat bind because of its configuration.

\subsection{Proxy Oracles}
\label{sec:formal-proxy}

A cheaper proxy could label a run from the shape of the use's destination argument (a raw path versus a descriptor or
  the descriptor-based mount API) or from an early rejection.
But the same shape can mean fixed for one contract and vulnerable for another, and a rejection shows only the fixes
  that reject.
We combine the two into the \emph{composite proxy}.

\noindent\textbf{A fix that is also a regression.}
In \texttt{runc}, commit \texttt{0ca91f44} both \textbf{fixes} CVE-2021-30465 and \textbf{introduces}
  CVE-2023-27561: its destination's shape (\texttt{string}$\to$\texttt{fd}) says \emph{fixed}, right for the first
  contract and wrong for the second, whose Race Specification labels it vulnerable (\S~\ref{sec:eval-revealed},
  Fig.~\ref{fig:one-commit-two-labels}).

\section{Chronos Design}
\label{sec:offense}

Chronos is a test controller driven by a YAML \emph{Race Specification} (\S~\ref{sec:dsl}).
It leaves the check $C$ to the target's own code and holds the use $U$ or a declared earlier checkpoint.
The oracle labels do not depend on lucky scheduling.

\subsection{Architecture: Race-as-Code}
\label{sec:architecture}

\begin{figure*}[t]
  \centering
  \includegraphics[width=\textwidth]{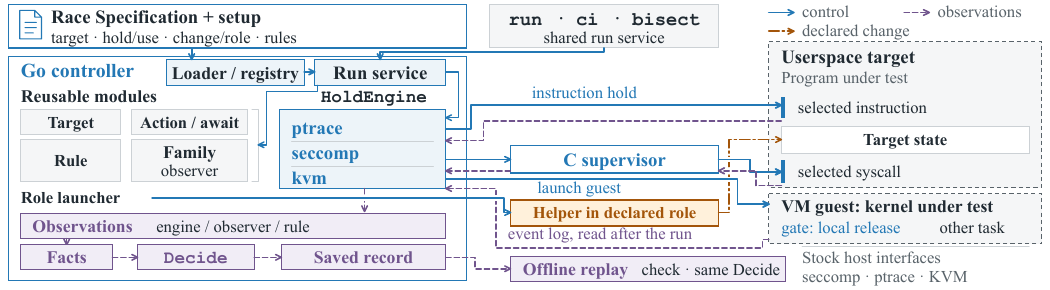}
  \caption{\textbf{Reusable modules feed one decision procedure.}
  The selected hold engine and modules supply observations that the shared core turns into facts.}
  \label{fig:chronos-framework-pair-architecture}
\end{figure*}

Chronos has a shared core and modules for hold engines, syscall families, attacker actions, targets and outcome
  rules (Fig.~\ref{fig:chronos-framework-pair-architecture}); each layer uses only lower layers.
The strict loader selects registered modules, and one run service coordinates setup, the intervention,
  observation and cleanup.
Only the seccomp engine uses the supervisor, written in C, which matches calls against a family-generated table
  (\S~\ref{sec:freeze-policy}).
The ptrace engine reaches a located userspace instruction directly; the kernel engine launches a guest whose
  gate releases locally, then reads the guest's event log after the run.
A per-build locator records the held instruction for each build under test and checks it against that build's code
  and debug information (its bytes, function and source line); the specification names only the function.
These observations and the rules' results feed the core's facts and its one decision procedure, \texttt{Decide}.
A family defines path arguments and effect observation (the \texttt{mount} family covers \texttt{mount},
  \texttt{open\_tree} and \texttt{move\_mount}); actions change and undo state, targets launch programs, and rules
  supply predicates over recorded observations or state. The family's observer runs in the controller.
Every driver (\texttt{run}, \texttt{ci}, \texttt{bisect} and \texttt{replay}) calls the same core.
A new syscall family, attacker action or kind of target therefore needs one new module and little or no change to the core
  (Table~\ref{tab:implementation}).
Most new races in a covered family need only a specification and the scripts that set up their scenarios.
Some have also needed a core change, such as a finer way to resolve names or to select which calls to hold;
  in our experience, such changes grew rarer as a family covered more races.
The syscall hold needs no kernel patch or kernel module; the hold at a kernel instruction loads its gate module only
  into the guest kernel under test.

\begin{table}[t]
  \caption{\textbf{A case generally added only its specification and setup scripts. Every other layer is shared.} Source lines
    per layer.}
  \label{tab:implementation}
  \centering
  \scriptsize
  \setlength{\tabcolsep}{3pt}
  \begin{tabularx}{\columnwidth}{@{}l Y r@{}}
    \toprule
    \textbf{Layer} & \textbf{Parts} & \textbf{Lines} \\
    \midrule
    Mechanisms & seccomp supervisor (C) & 3{,}282 \\
     & action helper and attacker launcher (C) & 830 \\
     & kernel gate (C), with its locator and guest init (Python, shell) & 883 \\
    Core (Go) & holds, identities, action runner, effect observation, decision, report and replay & 11{,}345 \\
    Modules (Go) & hold engines \texttt{ptrace}, \texttt{kvm} & 1{,}794 \\
     & syscall families \texttt{mount}, \texttt{open}, \texttt{stat}, \texttt{mkdir}, \texttt{create}, \texttt{unlink},
       \texttt{chmod} & 1{,}087 \\
     & actions \texttt{redirect}, \texttt{hide}, \texttt{run}, \texttt{await} & 2{,}141 \\
     & targets OCI runtime, command & 615 \\
     & outcome rules & 1{,}372 \\
    Drivers (Go) & CLI, bisection, specification set and module registry & 4{,}323 \\
    Support (Go) & process, identity, bounded I/O and UI & 939 \\
    Specifications & cases A--D, each & 6--11 \\
    Setup scripts (shell) & cases A--D, together & 275 \\
    \bottomrule
  \end{tabularx}
\end{table}

\subsection{Holding the Use}
\label{sec:freeze-policy}

\begin{figure*}[t]
  \centering
  \includegraphics[width=\textwidth]{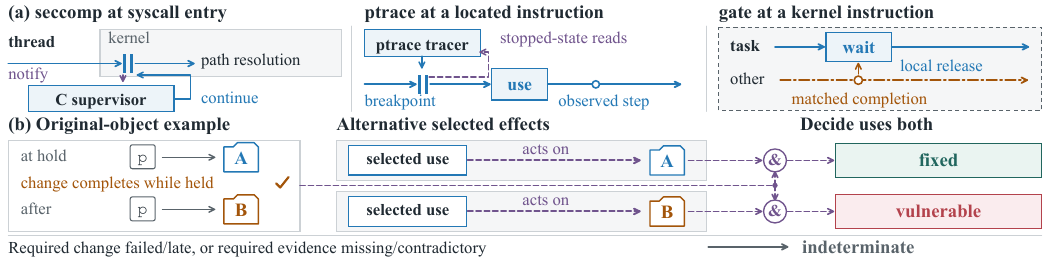}
  \caption{\textbf{Each engine stops the use, and \texttt{Decide} labels the run from the object the use acted on and
    the change's evidence.}
  In (b), a definitive label needs both, joined at each~\&; Table~\ref{tab:evidence-contracts}
  (Appendix~\ref{sec:appendix-evidence-contracts}) gives the other kinds of evidence.}
  \label{fig:chronos-framework-pair-evidence}
\end{figure*}

Chronos records the hold order, continuation and selected effect separately
  (Figs.~\ref{fig:race-window}(b) and~\ref{fig:chronos-framework-pair-evidence}).

For syscall holds, Chronos uses seccomp \texttt{USER\_NOTIF} (\chpt[ChAttack]{4} in Fig.~\ref{fig:intercept-map}).
For the syscalls of the held family, the filter returns \codebreak{SECCOMP_RET_USER_NOTIF},
  and the kernel blocks the calling thread at syscall entry until the Go controller requests release.
On release, the supervisor answers the kernel with \codebreak{SECCOMP_USER_NOTIF_FLAG_CONTINUE}.
We also tested two other stops in Chronos (\S~\ref{sec:eval-window}): a selective \texttt{ptrace} syscall stop
  (\chpt[ChAttack]{3}), which must follow every fork, clone and exec of the runtime and forward its signals, and a
  BPF stop (\chpt[ChAttack]{5}), which sends the thread \texttt{SIGSTOP} from the syscall tracepoint.

Before matching, the supervisor resolves descriptor arguments through the held thread, whether the argument is a name
  relative to a directory descriptor, an empty path with its descriptor, or a \texttt{/proc/<pid>/fd/N} path.
As a result, a fixed runtime that mounts or opens through a descriptor is held and attacked like a vulnerable one.

\subsection{Race Specification DSL}
\label{sec:dsl}

A Race Specification names the target and its setup scripts, the call to hold and, in its \texttt{attack} section,
  the attacker's change (Fig.~\ref{fig:spec-v3}).
The call to hold is a syscall family, a pattern for each path argument it constrains and, where needed, the calling
  program.
A specification may also name a later use than the held call (\texttt{crun}'s open search,
  \S~\ref{sec:eval-bisect}), an outcome rule where the object the use acted on does not decide the label, and
  bisection settings.
Every part the specification leaves out takes a default value, which Chronos records in the report.
The separate-use extension holds an earlier syscall checkpoint and observes a later deciding use; a checkpoint at a
  call that names no path, such as \texttt{umask}, declares the path as a literal. Execution between the checkpoint
  and the use continues.
The setup script also encodes the threat model in file permissions: the attacker can change whatever the script lets
  the attacker's user write, and whatever any user may write.

\begin{figure}[t]
\begin{lstlisting}[style=chronosyaml, basicstyle=\scriptsize\ttfamily\YAMLkeystyle, frame=none, xleftmargin=0pt, aboveskip=0pt, belowskip=0pt, postbreak=\mbox{$\hookrightarrow$\space}]
name: cve-2021-30465
target: {runtime: runc, bundle: build/bundle-30465}
setup: ./setup.sh
hold: {syscall: mount, destination: "*/mnt/shared/target"}
attack: {redirect: destination}
bisect: {repo: https://github.com/opencontainers/runc.git, from: v1.0.0-rc94, to: v1.0.0-rc95}
\end{lstlisting}
\caption{\textbf{Case A's Race Specification.}}
\Description{A six-line YAML file: the specification's name; the runc target and its bundle; the setup script; the
  mount call to hold, matched by its destination; the attack, which redirects that destination; and the bisection
  range in runc's repository.}
\label{fig:spec-v3}
\end{figure}

For the attacker's change, we provide a shared library of actions.
For instance, \texttt{redirect} exchanges one component of a held name with a symlink to a \emph{canary} in a single
  \texttt{renameat2(RENAME\_EXCHANGE)}, choosing by default the deepest component the attacker can change.
Chronos creates the canary itself, as a root-owned file or directory of the same type as the object the name pointed
  to, so the attacker cannot touch it.
A redirected use therefore acts with the target's authority on something the attacker could not reach.
The \texttt{hide} action moves a name out of the way and restores it after the call returns, \texttt{run} executes a
  helper command for any other change, and \texttt{await} waits for the program's own competing step, such as the free
  of \S~\ref{sec:eval-further}.
When the held argument is a descriptor, an action changes the path the descriptor resolved to at the hold.
Every action that makes a change runs as the case's attacker (\S~\ref{sec:realistic-cap}) and can change a name
  only through a path the attacker can reach in its own mount namespace (Table~\ref{tab:oracle-evidence}).
\begin{table*}[t]
  \caption{\textbf{Each fixed label rests on preservation under attack (A, D), a call never made (B) or the fix's own
    refusal (C).}}
  \label{tab:oracle-evidence}
  \centering
  \scriptsize
  \setlength{\tabcolsep}{4pt}
  \begin{tabularx}{\textwidth}{@{}p{0.11\textwidth}>{\hsize=0.8\hsize}Y>{\hsize=1.25\hsize}Y>{\hsize=0.85\hsize}Y>{\hsize=1.1\hsize}Y@{}}
    \toprule
    \textbf{Case} & \textbf{Hold} & \textbf{Change} & \textbf{Vulnerable} & \textbf{Fixed} \\
    \midrule
    \textbf{A}\newline CVE-2021-30465 & \texttt{mount()} by destination, after SecureJoin resolves it &
      \texttt{redirect}: the destination's last component exchanged with a symlink to a canary directory, by an
      unprivileged host user &
      redirected: the mount covers the canary &
      preserved: the descriptor-based mount is held, attacked, and covers the checked directory \\
    \textbf{B}\newline CVE-2025-23359 & the NVIDIA hook's \texttt{mount()}, by its \texttt{cuda/compat} source &
      \texttt{redirect}: the compat library exchanged with a symlink to a canary file, by an unprivileged host user &
      redirected: the bind exposes the canary &
      call never made: the shipped v1.17.4 hook makes its other mounts but never the compat bind \\
    \textbf{C}\newline CVE-2023-27561 & the procfs \texttt{mount()} &
      none: the image's \texttt{/proc} symlink into a shared tree is in place before the check &
      rule: procfs mounted at the image's path &
      refused: the fix rejects the \texttt{/proc} symlink with its own message, before the mount \\
    \textbf{D}\newline CVE-2025-31133 & runc's mask \texttt{mount()} of \texttt{/dev/null} over \texttt{/proc/kcore} &
      \texttt{hide}: \texttt{/dev/null} renamed away by root in a second container that shares its directory,
      restored after the call &
      rule: \texttt{/proc/kcore} unmasked &
      preserved: the descriptor-based mask mount is held, \texttt{/dev/null} hidden, and the mask still uses the real
      null device \\
    \bottomrule
  \end{tabularx}

  \vspace{2pt}
  {\scriptsize\raggedright Case D's two controls end indeterminate: held with no change, and with
    \texttt{/dev/zero} hidden instead of \texttt{/dev/null}.\par}
\end{table*}

Before execution, Chronos validates and normalizes the specification,
  and afterwards \texttt{chronos replay} recomputes the label from the report.
Replay checks the recorded observations and rule results for consistency.
The loader rejects unknown and duplicate keys, and families, actions and rules come from a closed set.
Every report carries a digest of the specification's canonical form, so each label traces to the exact specification
  behind it.
None of the four specifications needs a script of its own for the attacker's change (case A is in
  Fig.~\ref{fig:spec-v3}).

\subsection{Run Lifecycle}
\label{sec:lifecycle}

Each \texttt{chronos run} has nine phases: \chstep{1}~setup, \chstep{2}~readiness of any declared background process,
  \chstep{3}~arm, \chstep{4}~launch, \chstep{5}~hold, \chstep{6}~change, \chstep{7}~release, \chstep{8}~use and
  \chstep{9}~decide (Appendix~\ref{sec:appendix-supervisor} gives the seccomp-specific protocol for phases 5--9).
Setup completes the scenario; arm installs the hold that launch's target inherits.
Chronos creates a dedicated observer thread in the target's mount namespace, root and working directory.
The controller retains mount-table and root descriptors for checks after target exit.
After release, Chronos observes the use, undoes the change before the next intercepted family call, and writes the
  identity check, declared rule, label and evidence to the report.
For each declared use, the supervisor holds the first call that matches its syscall, its path patterns and the target
  executable; later calls are not evidence for it, so the specification must not match a benign earlier call.

\section{Deterministic Bisection and CI}
\label{sec:bisect}

A flaky exploit breaks the assumption behind \texttt{git bisect run} that each commit gets a reliable verdict.
If each natural trial hits with probability $q$, a vulnerable commit passes as fixed after $M$ misses
  with probability $(1-q)^M$.
Syzbot's cause bisection for the Linux kernel contained the culprit in 364 of 1{,}070 results up to March 2023;
  unstable reproduction or triage was the first error in 80\% of its failures, and races caused 12 of 20 sampled
  reproduction failures~\cite{gu2024bisection}.
Statistical bisectors model the noise: \emph{fixed retry} (Flake-Aware Culprit Finding's deflaked
  baseline~\cite{Henderson2023FlakeAware})
  calls a commit vulnerable at its first hit and fixed after $M$ misses, and \emph{Git Bayesect}~\cite{GitBayesect}
  keeps a posterior over the transition commit and stops at probability 0.95.
Deployed bisectors use fixed retry: syzbot with $M{=}10$~\cite{syzkaller2026bisect}, and V8's flake bisect with $M$
  set by doubling the repetitions until the test flakes~\cite{v8flakebisect}.

\begin{proposition}[Noisy-bisection bound]
\label{prop:bisect-bound}
Let $n \ge 2$, $0 < q \le 1$ and $0 \le \epsilon < 1 - 1/n$.
Let a range have one transition, at one of the $n$ commits after its older endpoint, chosen uniformly at random.
Let a bisector learn only from binary hit/miss outcomes of natural trials.
Conditional on the queried commit and the true transition, each trial's outcome is independent of earlier outcomes.
Its hit probability is at most $q$ at a vulnerable commit and zero at a fixed one.
The bisector picks each commit, and when to stop, from earlier outcomes.
To name the transition with error probability at most $\epsilon$, it runs on average at least
\[
  \frac{\log_2 n - h(\epsilon) - \epsilon\log_2(n-1)}{C(q)} \ \text{trials}.
\]
Here $h(p) = -p\log_2 p - (1-p)\log_2(1-p)$ is the binary entropy, with $h(0) = h(1) = 0$, and
\[
  C(q) = \max\nolimits_{0 \le x \le 1}\bigl[h(xq) - x\,h(q)\bigr].
\]
\end{proposition}

Appendix~\ref{sec:appendix-bayesect} sketches the proof, a classical channel-capacity argument with Fano's
  inequality~\cite{cover2006elements}.
In the CVE-2021-30465 Bayesect campaign, classified natural trials at vulnerable commits hit 20.4\% of the time.
We assume this rate is the ceiling $q$ at every vulnerable query.
Refusals and failed starts may carry further version information and lie outside this hit/miss bound.
A commit that fails before the race tells neither method where the transition is, so skips do not shrink $n$.
At $q$, a trial carries at most 0.116 bits; with a uniform prior over 19 commits, 95\% accuracy needs on average
  more than 32 trials, and six natural trials give at most 38\% accuracy.
As $q$ falls, $C(q)$ falls with it, and the bound grows without limit.
The bound covers every bisector that learns from natural hit/miss trials, whatever its inference.
A held trial is not such a trial: the declared change completes before the use, and a fixed label needs positive
  evidence.
An indeterminate trial moves no endpoint (Proposition~\ref{prop:bracket}).
\texttt{chronos bisect} runs one held trial per tested commit (\S~\ref{sec:timing-complete}).

\noindent\textbf{Labels and skips.}
At each midpoint, Chronos builds the revision in a pinned image of its toolchain and runs the same
  specification once, changing only the build under test.\label{sec:bisect-hermetic}
It maps the older endpoint's label to git's \emph{good}, the other to \emph{bad}, and an indeterminate run, timeout or
  failed build to \emph{skip}.

\noindent\textbf{Endpoint pre-validation.}
\texttt{chronos bisect} first labels both endpoints and aborts unless one is vulnerable and the other fixed; their
  order sets the fix or introduction direction, unlike plain \texttt{git bisect}, which trusts them.
Each search seeks one transition, so an introduction, a repair and a reintroduction need separate searches.

\begin{proposition}[Bracket]
\label{prop:bracket}
Fix the specification and the setup before a fix search on a linear range.
Suppose every revision in the searched range has one value under the declared contract, both definitive labels report
  that value soundly (\S~\ref{sec:timing-complete}), and the older endpoint is labeled vulnerable and the newer fixed.
Then some vulnerable-to-fixed transition always lies between the search's current vulnerable and fixed endpoints, and
  indeterminate runs never move them.
When the two endpoints are adjacent, the label changes between them: with one transition in the range, that is the
  transition; with several, it is a real one, though not necessarily the first or the last.
\end{proposition}
A search for an introduction is symmetric.
Appendix~\ref{sec:appendix-bracket} gives the proof.
\texttt{chronos bisect} checks that one endpoint is an ancestor of the other and records whether the range is
  linear or a merge graph; a range whose endpoints are not ancestral ends unverified.
On a linear range, the search identifies a label-changing boundary under Proposition~\ref{prop:bracket}'s premises.
When skips leave several candidates, Chronos reports that set.

\noindent\textbf{CI.}
\texttt{chronos ci} fails unless the label matches the one the caller expects.
An indeterminate run gets an exit code of its own, so CI can tell it apart from a pass or a failure.

\section{Evaluation}
\label{sec:evaluation}

We chose A--D in one syscall family, \texttt{mount}, to test whether its module, actions and rules carry over;
  later cases needed mainly a specification and small framework changes.
They span three kinds of fixed evidence, preservation (A, D), the fix's refusal (C) and a call never made (B),
  and B adds the NVIDIA hook beside runc.

Unless noted (Appendix~\ref{sec:appendix-baseline}), \texttt{runc} experiments ran on Ubuntu~24.04 with Linux~6.17;
  the searches, the release sweep, the reverts, the traces, Appendix~\ref{sec:eval-cases}'s two container races, the fault-injection audit
  and case B ran on Linux~7.0 (April 2026) cloud VMs.
Across the runs on cases A--D and Appendix~\ref{sec:eval-cases}'s two container races, apart from the BPF-stop and
  fixed-delay ablations, Chronos held the declared call in 1{,}999 runs and labeled 1{,}937 of them vulnerable or fixed.
The remaining 62, case D's two controls (Table~\ref{tab:oracle-evidence}) and 60 runs of six stress cases, such as
  signals and a killed supervisor, were designed to end indeterminate.
RQ1's effect-only checker reads 878 replayed records of A--D, and Appendix~\ref{sec:appendix-oracle-ablation}'s
  ablation 322 stored records of A--D.
Faults injected at each step of the supervisor's protocol, 23 in all, never produced a wrong definitive label
  (Appendix~\ref{sec:appendix-supervisor}).

RQ1 to RQ3 establish the oracle, on A--D in depth and on every other case under one bar: both builds labeled,
  the fixed side on positive evidence, and the fix isolated (Table~\ref{tab:cases}).
Every complete case meets it.
Every record behind these labels replays to its label.
RQ4 tests the fixes and what their records say.
\newcommand{\csY}{\chyes}
\newcommand{\csP}{\chpart}
\newcommand{\csN}{\chno}
\newcommand{\csNA}{--}
\newcommand{\csH}[1]{\rotatebox{90}{#1}}
\newcommand{\csG}[1]{\multicolumn{10}{@{}l}{\textit{#1}}\\}
\newcommand{\csMain}{\rowcolor{black!7}}
\begin{table*}[t]
  \caption{\textbf{Held trials label both builds of every complete case, and every record behind these labels replays
    to its label.} The depth columns show how far each case was taken; Table~\ref{tab:oracle-evidence} gives A--D's
    evidence in full.}
  \label{tab:cases}
  \centering
  \scriptsize
  \setlength{\tabcolsep}{2pt}
  \begin{tabularx}{\textwidth}{@{}>{\bfseries}l@{\hspace{3pt}}l l >{\hsize=0.96\hsize\raggedright\arraybackslash}X l
    >{\hsize=0.97\hsize\raggedright\arraybackslash}X >{\hsize=1.07\hsize\raggedright\arraybackslash}X
    crr@{}}
    \toprule
    & & & & & \multicolumn{2}{c}{\textbf{Bar}} & \multicolumn{3}{c@{}}{\textbf{Depth}} \\
    \cmidrule(lr){6-7}\cmidrule(l){8-10}
    & \textbf{Case} & \textbf{Program} & \textbf{Held at} & \textbf{Change} & \textbf{Vulnerable} & \textbf{Fixed} &
      \csH{Bisected} & \csH{Sweep} & \csH{Natural} \\
    \midrule
    \csG{Held at a syscall's entry}
    \csMain A & CVE-2021-30465 & runc & \texttt{mount} by destination & \texttt{redirect} & rc94, 100/100: redirected &
      rc95, 50/50: preserved & \csY & 75 & 603/2{,}000 \\
    \csMain B & CVE-2025-23359 & NVIDIA hook & the compat \texttt{mount} & \texttt{redirect} &
      1.17.3, 10/10: redirected & 1.17.4, 10/10: call never made & \csY & 15 & 31/200$^{a}$ \\
    \csMain C & CVE-2023-27561 & runc & the procfs \texttt{mount} & none & parent, 20/20: rule & fix, 20/20: refused &
      \csY & 75 & \csNA$^{b}$ \\
    \csMain D & CVE-2025-31133 & runc & the mask \texttt{mount} & \texttt{hide} & parent, 20/20: rule &
      fix, 20/20: preserved & \csY & 75 & 8/200 \\
    & CVE-2024-45310 & runc & \texttt{mkdir}, \texttt{create} & \texttt{redirect} & v1.1.13, 10/10: redirected &
      v1.1.14, 10/10: preserved & \csY & 73 & 15/100$^{c}$ \\
    & CVE-2021-35938 & rpm & \texttt{chmod} on upgrade & \texttt{redirect} & 4.18.0$^{d}$, 10/10: redirected &
      4.18.1, 10/10: preserved & \csY & 13 & 17/100 \\
    & CVE-2013-4235 & shadow & \texttt{unlink} by \texttt{userdel -r} & \texttt{redirect} & 4.12.1, 10/10: redirected &
      4.12.2, 10/10: preserved & \csY & 41 & 16/100$^{e}$ \\
    & socket race~\cite{chronySocketReport2026} & chrony & chronyc's socket \texttt{chmod} & \texttt{redirect} & 4.9, 10/10: redirected &
      4.9 + fix~\cite{chronySocketFix2026}, 10/10: preserved & \csY & & 1/1{,}200 \\
    & CVE-2019-19921 & runc & \texttt{mount} of \texttt{/proc/sys} & \texttt{redirect} & parent, 10/10: redirected &
      fix, 10/10: refused & & & 15/200 \\
    & CVE-2021-25741 & kubelet & the subPath \texttt{mount} & \texttt{redirect} & parent, 10/10: redirected &
      fix, 10/10: preserved & \csY & 37 & 25/200 \\
    & CVE-2018-15664 & Docker & \texttt{open} in \texttt{docker cp} & \texttt{redirect} & parent, 10/10: redirected &
      fix, 10/10: rule & \csY & 14 & 44/200 \\
    & SUSE bug 1241543 & Cyrus & \texttt{umask} before \texttt{chmod} & \texttt{redirect} & 3.12.0, 10/10: redirected &
      3.12.1, 10/10: call never made & \csY & 121 & 0/200 \\
    & CVE-2026-43619 & rsync & \texttt{chmod}, \texttt{rmdir}, \texttt{mkdir}, \texttt{unlink} & \texttt{redirect} &
      3.4.2, 10/10: redirected & 3.4.3, 10/10: preserved & \csY & 31 & 0/200 \\
    & CVE-2025-5054 & apport & the forwarding \texttt{open} & \texttt{redirect} & parent, 10/10: rule &
      fix, 10/10: call never made & \csY & 25$^{f}$ & 200/200$^{g}$ \\
    & CVE-2025-52565 & runc & the console \texttt{mount} & \texttt{redirect} & 1.2.7, 10/10: redirected &
      1.2.8, 10/10: preserved & \csY & 41 & 200/200$^{h}$ \\
    & CVE-2025-6020 & Linux-PAM$^{i}$ & the polydir \texttt{mkdir} & \texttt{redirect} & 1.7.0, 10/10: redirected &
      1.7.0 + fix, 10/10: preserved & \csNA & 40 & 0/200 \\
    & CVE-2021-23240 & sudo$^{j}$ & sudoedit's \texttt{chown} & \texttt{redirect} & 1.9.4p2, 10/10: redirected &
      1.9.4p2 + fix, 10/10: preserved & \csNA & 46 & 9/200 \\
    \midrule
    \csG{Held at a userspace instruction (\texttt{ptrace})}
    & CVE-2014-0226 & Apache & \texttt{mod\_status}'s copy & a worker's update & 2.4.9, 10/10: ASan report &
      2.4.10, 10/10: rule & \csY & 50 & 11/100 \\
    & CVE-2021-4207 & QEMU & the cursor's second read & the guest's write & parent$^{k}$, 10/10: ASan report &
      fix, 10/10: rule & \csY & 4 & 200/200$^{l}$ \\
    \midrule
    \csG{Held at a kernel instruction (the kernel gate)}
    & CVE-2022-49919 & Linux & the freed chain's read & the free & parent, 10/10: KASAN report & fix, 10/10: call never made &
      \csY & 4 & 20/20 \\
    & iommufd \texttt{6f9c4d8c} & Linux & the I/O address space's read & the free & parent, 10/10: KASAN report &
      fix, 6/10$^{m}$: call never made & \csY & 10 & 10/20 \\
    \midrule
    \csG{One-sided: the fixed side cannot be labeled$^{n}$}
    & \texttt{l2cap\_chan\_close} & Linux & the 6LoWPAN read & the free & 3/30: KASAN report & fix, 10/10: indeterminate &
      \csNA & \csNA & \\
    & \texttt{sk\_msg\_recvmsg} & Linux & the held read & the free & 1/21: KASAN report &
      \texttt{c010995b}, 10/10: indeterminate & \csNA & \csNA & \\
    \bottomrule
  \end{tabularx}

  \vspace{3pt}
  {\scriptsize\raggedright \csY{} done; blank: not measured; \csNA{} not applicable. Shaded: the main cases. ``+ fix'':
    the last vulnerable release with only the fix applied.\par
    \smallskip
    Counts are trials labeled so, of trials run. Parent, fix: the fix's exact parent and the fix; C's and D's come from
    RQ1's reverts. Bar, required of every complete case and met by each: both builds labeled, the fixed side on
    positive evidence, and the fix isolated (the fix alone labeled both ways, reverted or applied by itself).
    Bisected: each search is a row of Table~\ref{tab:bisect-summary}; a ``+ fix'' pair differs by the fix alone, so it
    has no range to search (\csNA), except chrony's, whose fix has two commits. Sweep: releases of the case's program that the release sweep labeled vulnerable or
    fixed; A, C, D and CVE-2024-45310 also ran on \texttt{crun}'s and \texttt{youki}'s releases
    (Table~\ref{tab:release-sweep}).
    Natural: the natural rate, natural runs that reached the effect at a vulnerable build, of runs made.\par
    \smallskip
    \noindent\begin{minipage}[t]{0.4875\linewidth}
    \chNote{a}{B's other runs: in 117 the exploit lost the race, and in 52 the swap came before the hook's check,
      which refused the symlinked source.}
    \chNote{b}{C's change is in place before the check, so there is no window to miss.}
    \chNote{c}{CVE-2024-45310's natural runs had one victim per launch.}
    \chNote{d}{4.18.0 shipped this CVE's fix, which missed the upgrade path (Fig.~\ref{fig:fix-lifecycle}; Table~\ref{tab:findings}).}
    \chNote{e}{shadow's natural runs used the held runs' directory layout.}
    \chNote{f}{apport's other 77 tags, all before 2.21.0, make no held call: 11 connect to the forwarding socket by
      pathname, which the specification does not hold, and 66 have no forwarding socket.}
    \chNote{g}{apport's natural runs started the attacker's socket swap a uniformly drawn 0 to 60~ms after launching apport; at those delays the swap lands before apport's \texttt{open}.}
    \end{minipage}\hfill
    \begin{minipage}[t]{0.4875\linewidth}
    \chNote{h}{CVE-2025-52565's attack at 1.2.7 is a static symlink and needs no race.}
    \chNote{i}{Linux-PAM's attack works on a stock systemd host from 1.1.6 on, and before that needs the session's
      mounts kept private.}
    \chNote{j}{sudo's runs set SELinux permissive and \texttt{fs.protected\_symlinks=0}, the advisory's own
      preconditions.}
    \chNote{k}{QEMU's parent is 7.0.0 with the fix reverted, and its fix is 7.0.0 itself.}
    \chNote{l}{QEMU's natural runs each queue 20{,}000 cursor commands.}
    \chNote{m}{The iommufd fix's other four trials were indeterminate, with no free observed during the hold.}
    \chNote{n}{Each fix keeps the read and makes it safe by ordering or a lifetime reference; a fixed label needs
      ordering evidence that the engines do not give, so the bar, a search and a sweep do not apply (\csNA).}
    \end{minipage}\par}
\end{table*}

\subsection{RQ1: Is One Held Trial Decisive?}
\label{sec:eval-window}

Yes.

\paragraph{Held and natural runs}
For CVE-2021-30465, the exploit without the hold exposed host-root content in 603 of 2{,}000 attempts (30\%).
In the rest, runc refused the swapped mount in 685, and the exploit lost the race in 712, 315 of them with the
  final check off the destination.
The 685 were SecureJoin \texttt{readlink} errors at the vulnerable parent, so a refusal alone is not evidence of a
  fix.
With the hold, the swap redirected the mount to a canary directory in all 100 runs of the vulnerable version.
In all 50 runs of the fixed version, the held mount reached the original destination through its descriptor, and
  Chronos labeled it fixed.
The interval from runc's final SecureJoin check of the destination to the kernel's \texttt{path\_mount} entry, just
  after the kernel resolves it, is an upper bound on $W$; over the 1{,}000 runs whose final check was on the destination
  itself, its median was $27.9\,\mu$s (p5--p95: $14.4$--$53.4\,\mu$s).

\begin{table}[t]
  \caption{\textbf{Without the hold, the CVE-2021-30465 exploit succeeds in 8--94\% of attempts or launches
    on vulnerable builds across setups and batches. With the hold, the swap redirects the mount to a canary in all 100
    runs.}}
  \label{tab:natural-rates}
  \centering
  \scriptsize
  \setlength{\tabcolsep}{2pt}
  \begin{tabular*}{\columnwidth}{@{\extracolsep{\fill}}lccrr@{}}
    \toprule
    \textbf{Setup} & \textbf{Victims} & \textbf{Linux} & \textbf{Hits} & \textbf{Rate} \\
    \midrule
    Window campaign (RQ1) & 1 & 6.17 & 603/2{,}000 & 30\% \\
    Fixed-retry campaign (RQ3) & 1 & 7.0 & 124/657 & 19\% \\
    Git Bayesect campaign (App.~A) & 1 & 7.0 & 182/894 & 20\% \\
    4 swappers, local (App.~A)$^{*}$ & 1--32 & 6.17 & 8--22/100 & 8--22\% \\
    4 swappers, cloud VM (App.~A)$^{*}$ & 4, 19 & 7.0 & 70, 84/100 & 70, 84\% \\
    4 swappers, amplified campaign (RQ3)$^{*}$ & 19 & 7.0 & 94/100 & 94\% \\
    \quad fixed control$^{*}$ & 19 & 7.0 & 0/50 & 0\% \\
    \midrule
    Chronos, held (RQ1) & 1 & 7.0 & 100/100 & 100\% \\
    \bottomrule
  \end{tabular*}

  \vspace{3pt}
  {\scriptsize\raggedright One swapper unless stated otherwise. Bisection rows count trials at vulnerable commits.
    $^{*}$Per launch: a launch starts its victims at once and succeeds when any escapes.\par}
\end{table}

\paragraph{Other stops}
The \texttt{ptrace} syscall stop held and labeled vulnerable all 100 vulnerable runs of cases A and D at about the
  same cost per stop (158\,$\mu$s for \texttt{ptrace} against 161\,$\mu$s for \texttt{USER\_NOTIF}), and in six stress cases neither mislabeled a run.
On the case-A endpoints with one swap per trial (Table~\ref{tab:hold-mechanisms}), the BPF stop took effect only
  after the mount committed, so it needs an action-specific undo; a fixed delay gave the effect only when it
  outlasted the swap (0.70\,ms median, 0.94\,ms maximum).
Release on completion needs no delay tuned to how long the helper takes.

\begin{table}[t]
  \caption{\textbf{Release on completion gave the effect in every vulnerable trial, an asynchronous stop in none, and a
    fixed delay only when it outlasted the swap.} Stops for one swap on the case-A endpoints.}
  \label{tab:hold-mechanisms}
  \centering
  \scriptsize
  \setlength{\tabcolsep}{4pt}
  \begin{tabular*}{\columnwidth}{@{\extracolsep{\fill}}llrr@{}}
    \toprule
    \textbf{Stop} & \textbf{Release} & \shortstack[r]{\textbf{Effect,}\\\textbf{vulnerable}} &
      \shortstack[r]{\textbf{Effect,}\\\textbf{fixed}} \\
    \midrule
    BPF \texttt{SIGSTOP} & -- & 0/200 & 0/100 \\
    \texttt{USER\_NOTIF} & short fixed delay & 0/200 & 0/100 \\
    \texttt{USER\_NOTIF} & long fixed delay & 200/200 & 0/100 \\
    \texttt{ptrace} & on completion & 20/20 & \\
    \texttt{USER\_NOTIF} (Chronos) & on completion & 20/20 & 0/20 \\
    \bottomrule
  \end{tabular*}

  \vspace{3pt}
  {\scriptsize\raggedright Effect: trials in which the swap changed the completed mount. Each fixed-delay row gives the count at each of its delays: short $0$, $10$ and
    $100\,\mu$s; long $1$, $10$ and $100$\,ms. Blank: not measured.\par}
\end{table}

\paragraph{Reverted fixes}
In our runs, undoing a fix never produced a fixed label.
We reverted the fixes of cases A, C and D in 13 ways, from the fix's exact parent to versions that bring the bug back
  through a different call than the held one or emit the fix's refusal message without refusing, 20 runs each.
The 160 runs of the eight reverts that kept the bug at the held call were labeled vulnerable; the other five moved it
  to a call the specification does not hold, and their 100 runs, holding no call, were indeterminate.
All 60 runs of the three fixes themselves were labeled fixed.
Case B's fixed label rests on a call never made: in its default configuration, the fixed release never made the
  held call in 10 of 10 runs, and its container's mount table showed no compat bind; with the vulnerable hook's
  program renamed so that the hold could not match it, all 10 runs were indeterminate and none fixed
  (Appendix~\ref{sec:eval-cases}).

\paragraph{A cheaper proxy}
\label{sec:eval-proxy}
The composite proxy of \S~\ref{sec:formal-proxy}, run on \texttt{runc}'s ranges without \texttt{USER\_NOTIF}, gives
  no label at CVE-2023-27561's vulnerable endpoint, where the destination's form says fixed while the procfs mount
  says vulnerable, so it cannot bisect either of that CVE's ranges.
On both CVE-2021-30465 ranges, where form and security change in the same commit, the proxy matches Chronos
  at all four and ten midpoints, respectively.
For CVE-2025-31133 it returns \texttt{7ea4eb02}, a later commit that passes the destination as a descriptor, because
  the proxy reads the destination and the fix \texttt{11cf22d1} changes the source.
Keyed to the source, it names \texttt{11cf22d1} as Chronos does: the right key depends on which argument the fix
  changes.

\begin{table}[t]
  \caption{\textbf{Checkers that read only the effect labeled runs of vulnerable builds fixed when the held use showed
    no effect; Chronos labeled none of them fixed.}}
  \label{tab:effect-only}
  \centering
  \scriptsize
  \setlength{\tabcolsep}{3pt}
  \begin{tabular*}{\columnwidth}{@{\extracolsep{\fill}}lrrrr@{}}
    \toprule
    & & \multicolumn{2}{c}{\textbf{Effect only}} & \\
    \cmidrule(lr){3-4}
    \textbf{Records} & \textbf{Runs} & \shortstack[r]{\textbf{missing}\\\textbf{effect}\\\textbf{as fixed}} &
      \shortstack[r]{\textbf{missing}\\\textbf{effect}\\\textbf{as indet.}} & \textbf{Chronos} \\
    \midrule
    A--D, replayed & 878 & 82 & 42 & 0 \\
    Other stops and controls, A and C & 2{,}162 & 800 & 800 & 0 \\
    \texttt{sk\_msg\_recvmsg} & 21 & 20 & 20 & 0 \\
    \midrule
    Total & 3{,}061 & 902 & 862 & 0 \\
    \bottomrule
  \end{tabular*}

  \vspace{3pt}
  {\scriptsize\raggedright Cells: runs of a vulnerable build labeled fixed. Both checkers label fixed a run whose
    effect was observed absent; they differ on a missing effect, one never observed, which the first takes for fixed
    and the second for indeterminate. A--D's 878 records: 401 endpoint, 60 revert, 200 mutant, 97 midpoint and 120
    control runs.\par}
\end{table}

\paragraph{Reading only the effect}
A checker that reads only the use's effect agrees with Chronos on every record in which the hold and the change
  completed; they differ exactly where a check of Chronos's fails: the use was held, the change completed before
  release, and the effect is the held call's (\S~\ref{sec:freeze-soundness}).
On 878 replayed records of cases A--D (endpoints, reverts, mutants, the searches' midpoints and controls;
  Table~\ref{tab:effect-only}), the checker
  that takes a missing effect for fixed labeled 82 runs fixed whose version other held runs show vulnerable.
Taking a
  missing effect for indeterminate instead still left 42, and Chronos labeled none.
On 2{,}162 trials of alternative stops and controls for cases A and C, both checkers labeled fixed 800 trials of case A's vulnerable version in which the change had not
  acted before the use: the stop came after the mount, or the call was released before the helper finished.
On 21 trials of a use-after-free in the kernel's \texttt{sk\_msg\_recvmsg}, a one-sided kernel case of
  \S~\ref{sec:eval-further}, Chronos labeled one vulnerable, with the sanitizer's report on the held
  object, and 20 indeterminate, with no matching free observed during the hold; both checkers labeled those 20 fixed,
  although the kernel is vulnerable.

In ten of Table~\ref{tab:bisect-summary}'s searches (A--D's eight, \texttt{crun}'s held-mount search and
  \texttt{youki}'s), the checker that takes a missing effect for fixed followed Chronos in all ten.
  The one that takes it for indeterminate could not start case B's three, whose fixed endpoints make no held call.
Both checkers correctly labeled 40 records of two case-C mutants vulnerable; Chronos labeled them indeterminate,
  because procfs reached the forbidden mount through a call the hold did not see.
For CVE-2023-27561, a final-state check alone told the versions apart: the procfs mount appeared in 10 of 10
  vulnerable runs and in none of 10 fixed runs.

\subsection{RQ2: Does It Carry Over to Other Hold Engines and Bug Classes?}
\label{sec:eval-further}

Our primary evaluation concerns TOCTOU races held at syscall entry.
Table~\ref{tab:cases} gives all twenty-three named cases, A--D included, grouped by hold site.
Five further cases, each under its own attacker (\S~\ref{sec:threat}) and blocking premise
  (\S~\ref{sec:timing-complete}), test
  whether the specification format and the oracle carry past the four main cases (Table~\ref{tab:cases}):
  one in container setup outside the \texttt{mount} family, and four beyond container setup; three are held at a
  syscall's entry (CVE-2024-45310, rpm and shadow), one at a userspace instruction through a \texttt{ptrace} hardware breakpoint (Apache), and one at a
  kernel instruction through the kernel gate, in a KVM guest (CVE-2022-49919).
The instruction cases demonstrate the same decision procedure with userspace and kernel holds.
Every label uses the evidence rules of \S~\ref{sec:freeze-soundness} unchanged: each hold engine only reports
  facts (the hold reached in this build, the change complete while the use was held, what the use acted on), and
  one decision procedure labels them, reading a sanitizer's report on the held object as a declared rule and the
  fixed kernel's missing read as a call never made.
Reverting only the fix, at the fixed release or at its commit, made each of the three syscall-entry cases vulnerable
  again in 10/10 held trials (per call for runc).
The five further cases' natural rates are 11--100\% (Table~\ref{tab:cases}); for them, the hold's main benefit is
  the fixed-side label, which rests on positive evidence that an unheld miss cannot provide.
Two add families for directory entries: runc's CVE-2024-45310 creates a file or directory through a swapped parent
  (\texttt{mkdir}, \texttt{create}), and shadow's CVE-2013-4235 lets \texttt{userdel -r} delete outside the home
  directory (\texttt{unlink})~\cite{runcCve202445310,shadowCve20134235}.
rpm extended the \texttt{chmod} family with a small core change, and Apache and
  the kernel race use the instruction-hold engines of Table~\ref{tab:implementation}.

\paragraph{A userspace and a kernel hold}
In Apache httpd's CVE-2014-0226, \texttt{mod\_status} measures another worker's request line in the shared
  scoreboard, then copies it~\cite{apacheCve20140226}.
A \texttt{ptrace} hardware breakpoint holds the reader between the measurement and the copy while another worker
  updates the entry through a real request.
On release, 2.4.9 built with AddressSanitizer reported a one-byte write out of bounds in \texttt{ap\_escape\_logitem},
  and 2.4.10 copied its checked snapshot.
Without the hold, every hit in runs of 1{,}000 request rounds was a one-byte over-read from the same double read,
  never this write.
Ten held trials bisected the history to the repair merge.
For CVE-2022-49919, an \texttt{nf\_tables} use-after-free with a syzbot reproducer~\cite{meerkat2026,syzbotNftables93},
  the kernel gate, in a two-vCPU KVM guest running a KASAN debug kernel, parks the sleepable worker before its read
  of the freed chain while the reproducer frees the
  chain on the other vCPU; the action \texttt{await} waits for that free.
The vulnerable kernel gave KASAN's report on the held object in 10 of 10 boots, and the fixed kernel completed the
  same schedule in 10 of 10, with source and compiled-code checks showing that the selected read was absent.
Replay reproduced each label; one trial on each side of the introducing commit agreed with Meerkat's boundary.

\paragraph{Four more syscall-entry cases}
For kubelet, \texttt{runc} and Docker, the labels that follow are at their releases (three runs per endpoint,
  Appendix~\ref{sec:eval-cases}); Table~\ref{tab:cases} counts each fix's exact parent and the fix.
Held at kubelet's subPath bind mount (CVE-2021-25741), v1.21.4 exposed the canary and v1.21.5, binding through a
  descriptor path, the original directory.
In \texttt{runc}, rc9's mount of \texttt{/proc/sys} (CVE-2019-19921) was redirected, and rc10 refused it.
Cyrus IMAP 3.12.0's permission-changing \texttt{chmod} after a held \texttt{umask} checkpoint was redirected, and
  3.12.1 made no matching \texttt{chmod}.
Without the hold, none of 200 natural runs on 3.12.0 reached the effect; all 10 held trials did.
Docker 18.09.6's copy was redirected at its held \texttt{open}, and 18.09.7 was labeled fixed by its declared
  confinement rule.

\paragraph{Scope beyond A--D}
Beyond A--D, seventeen cases have definitive labels at both endpoints (Table~\ref{tab:cases}).
The one-sided kernel cases \texttt{l2cap\_chan\_close} and \texttt{sk\_msg\_recvmsg} have vulnerable labels only.
Of 30 held \texttt{l2cap\_chan\_close} trials at the fix's parent, 3 recorded a matching free and KASAN at the held
  read, and 27 were indeterminate.
The \texttt{sk\_msg\_recvmsg} case has one vulnerable and twenty indeterminate trials (\S~\ref{sec:eval-proxy}).
With an active action the attacker makes the change during the hold, so one trial labels the build or ends
  indeterminate; with \texttt{await} the change is the target's own step, so a trial may abstain, and a repeated
  trial is as sound as the first.

\subsection{RQ3: Does One Held Trial per Commit Suffice for Bisection and CI?}
\label{sec:eval-bisect}

Yes, in every search: all 27 bisections over real Git histories (Table~\ref{tab:bisect-summary}) converged to the
  boundary commit known from upstream.
The eight runc and NVIDIA searches' 47 midpoint labels, one trial each, were all definitive and on the expected side,
  with no skips.
Of the 17 midpoints labeled fixed, 6 rest on preservation under attack, 8 on a declared refusal and 3 on a
  declared call never made.
With the opt-in compat mode, v1.18.0 was labeled vulnerable (Table~\ref{tab:bisect-summary}).

\paragraph{Repairs across runtimes}
With its runtime and repository changed, case A's specification bisected \texttt{crun}'s and \texttt{youki}'s
  histories to one repair commit each, both labeled fixed by preservation under attack.
\texttt{crun}'s \texttt{afeb0654} mounts through \codebreak{/proc/self/fd} paths that it opens by name and leaves
  unchecked until \texttt{05161684}, which a second search, holding that open, named.
The fixed label for \texttt{afeb0654} therefore covers only the held mount.
\texttt{youki}'s \texttt{5886c910} moves the held call from \texttt{mount(2)} on a path to \texttt{move\_mount(2)} on
  a descriptor.

\begin{table}[t]
  \caption{\textbf{All 27 bisections and the scans of rpm and chrony found their boundary.} None skipped a midpoint.}
  \label{tab:bisect-summary}
  \centering
  \scriptsize
  \setlength{\tabcolsep}{2pt}
  \begin{tabular*}{\columnwidth}{@{\extracolsep{\fill}}lrrrl@{}}
    \toprule
    \textbf{Search} & \textbf{Commits} & \textbf{Midpoints} & \textbf{Time} & \textbf{Found} \\
    \midrule
    \tabgroup{5}{A: CVE-2021-30465 (runc); fixed by preservation}
    fix, narrow range & 19 & 4 & 9\,s & \texttt{0ca91f44} \\
    fix, wide range & 1{,}267 & 10 & 95\,s & \texttt{0ca91f44} \\
    \tabgroup{5}{B: CVE-2025-23359 (NVIDIA); fixed by a call never made or refusal}
    fix & 26 & 5 & 8\,s & \texttt{f2b3e8d3} \\
    public \texttt{--no-cntlibs} flag & 3 & 2 & 36\,s & \texttt{7e3642db} \\
    opt-in compat mode & 213 & 7 & 127\,s & \texttt{8ed5824a} \\
    \tabgroup{5}{C: CVE-2023-27561 (runc); fixed by refusal}
    introduction & 651 & 9 & 32\,s & \texttt{0ca91f44} \\
    fix & 27 & 5 & 11\,s & \texttt{0abab45c} \\
    \tabgroup{5}{D: CVE-2025-31133 (runc); fixed by preservation}
    fix & 32 & 5 & 17\,s & \texttt{11cf22d1} \\
    \tabgroup{5}{Case A's attack across runtimes; fixed by preservation}
    \texttt{crun}, the held mount & 67 & 6 & 123\,s & \texttt{afeb0654} \\
    \texttt{crun}, the held open (refused) & 43 & 6 & 131\,s & \texttt{05161684} \\
    \texttt{youki} fix & 11 & 4 & 285\,s & \texttt{5886c910} \\
    \tabgroup{5}{CVE-2024-45310's attack across runtimes; fixed by preservation}
    \texttt{crun}, directory & 67 & 6 & 117\,s & \texttt{8e580bcb} \\
    \texttt{crun}, file & 67 & 6 & 118\,s & \texttt{8e580bcb} \\
    \texttt{youki}, directory & 11 & 4 & 243\,s & \texttt{5886c910} \\
    \texttt{youki}, file & 11 & 4 & 233\,s & \texttt{5886c910} \\
    \tabgroup{5}{Further cases (Table~\ref{tab:cases}): the fix}
    CVE-2024-45310 (\texttt{runc}) & 24 & 5 & 21\,s & \texttt{f0b652ea} \\
    CVE-2025-52565 (\texttt{runc}) & 32 & 5 & 29\,s & \texttt{0c3e7c1b} \\
    shadow, the held use$^{a}$ & 8 & 3 & 84\,s & \texttt{f6f8bcd2} \\
    Apache httpd & 415 & 8 & 193\,s & \texttt{4bbf0508} \\
    kubelet & 42 & 6 & 4\,s & \texttt{d2609d19} \\
    Docker & 108 & 7 & 63\,s & \texttt{61e04590} \\
    rsync & 34 & 6 & 81\,s & \texttt{862fe4ee} \\
    Cyrus & 31 & 5 & 67\,s & \texttt{20207b4a} \\
    apport & 10 & 4 & 13\,s & \texttt{907a2b7f} \\
    QEMU & 5 & 2 & 144\,s & \texttt{9569f5cb} \\
    CVE-2022-49919 (\texttt{nf\_tables}) & 4{,}980 & 12 & 6{,}002\,s & \texttt{26b5934f} \\
    the iommufd fix & 3 & 1 & 219\,s & \texttt{6f9c4d8c} \\
    \tabgroup{5}{rpm: a scan, one held trial per commit}
    fix & 80 & -- & 418\,s & \texttt{8b5cce9a} \\
    \tabgroup{5}{chrony: a scan of the proposed fix's two commits, three held trials each}
    proposed fix & 2 & -- & & \texttt{04a23627} \\
    \bottomrule
  \end{tabular*}

  \vspace{3pt}
  {\scriptsize\raggedright Commits: those after the older endpoint, as \texttt{git} counts the range. Midpoints
    exclude two validated endpoints. The bisections executed $\lfloor\log_2 n\rfloor$ or
    $\lceil\log_2 n\rceil$ midpoint trials for $n$ commits. \texttt{crun}'s setup marks its bind mounts \texttt{rprivate}, so every commit in its ranges starts.
    Time: the search's duration after endpoint validation, including midpoint builds; blank where the records do
    not report it.
    $^{a}$The repair of \texttt{remove\_tree}, the held use, one of 4.12.2's seven repair commits.\par}
\end{table}

\paragraph{Natural walks}
We bisected without the hold as a practitioner would: each attack's one-victim setup, one natural launch per
  trial, and two policies, fixed retry with a per-walk $M$ (Table~\ref{tab:amplified-baseline}) and Git Bayesect
  v1.2's 0.95 posterior stopping rule.
Their cost followed the race's rate.
At case A's 21.5\%, both found \texttt{runc}'s fix at medians of 36 and 165 launches over 19
  commits, and at case D's 4.0\%, at 475 and 398; Chronos's searches used six and seven held trials.
On \texttt{crun}, at 0.10\%, capped walks stopped after 726--900 launches without the fix, \texttt{05161684}.
  With $M{=}10{,}921$, fixed retry found it in one of five walks, after 39{,}159 launches, and took
  \texttt{afeb0654} for it in the other four; Git Bayesect named it in none of five, four of them stopping at its
  40{,}000-launch limit. Chronos's searches used 8 held trials for the mount and 8 for the open.
With the discoverer's 19 simultaneous victims and four swappers, case A's endpoint hit in 94/100
  launches (Wilson 95\% 87.5--97.2\%) and 0/50 at the fix (upper bound 7.14\%), and both policies found the fix
  in five of five walks, at medians of 8 and 12 launches, with no skipped or unclassified launch.
This amplification is specific to the attack, which needs many victims that share the attacker's volume, and it still yields a
  rate: refusals occurred in 56/100 vulnerable and all 50 fixed launches, so neither a refusal nor a miss gives a
  positive fixed label.

\begin{table}[t]
  \caption{\textbf{With one victim, natural bisection cost more as the race got rarer, and at \texttt{crun}'s rate 1 of 20
    walks found the fix; the discoverer's amplified attack made case A's walks cheap.} Chronos needed 6 to 16 trials.}
  \label{tab:amplified-baseline}
  \centering
  \scriptsize
  \setlength{\tabcolsep}{2pt}
  \begin{tabular*}{\columnwidth}{@{\extracolsep{\fill}}lrrrrr@{}}
    \toprule
    \textbf{Search (commits)} & \shortstack[r]{\textbf{Rate}\\\textbf{(\%)}} & \textbf{Bound} &
      \shortstack[r]{\textbf{Fixed retry}\\\textbf{right, median}} & \shortstack[r]{\textbf{Bayesect}\\\textbf{right, median}} &
      \textbf{Chronos} \\
    \midrule
    \multicolumn{6}{@{}l}{\textit{One victim per launch}} \\
    \texttt{runc} A (19)   & 21.5 & 30$^{a}$ & 5/5, 36  & 5/5, 165  & 6 \\
    \texttt{runc} A (1{,}267) & 21.5 & 77$^{a}$ & 5/5, 68 & 4/5, 235 & 12 \\
    \texttt{runc} D (32)   & 4.0  & 207   & 5/5, 475  & 5/5, 398  & 7 \\
    \texttt{crun} A (67), capped & 0.10 & 10{,}317 & 0/5$^{b}$, 818 & 0/5$^{c}$, 800 & 16$^{d}$ \\
    \texttt{crun} A (67), uncapped & 0.10 & 10{,}317 & 1/5, 27{,}350 & 0/5$^{c}$, 40{,}000 & 16$^{d}$ \\
    \multicolumn{6}{@{}l}{\textit{19 victims and four swappers per launch}} \\
    \texttt{runc} A (19)   & 94   & 4     & 5/5, 8    & 5/5, 12   & 6 \\
    \bottomrule
  \end{tabular*}

  \vspace{3pt}
  {\scriptsize\raggedright Rate: hits at the vulnerable endpoint in the walks' own calibration (43/200, 8/200,
    2/2{,}000, 94/100; none at the fixed one). Bound: Proposition~\ref{prop:bisect-bound}'s, at that rate
    (Table~\ref{tab:bisect-races}, Appendix~\ref{sec:appendix-baseline}, has the intervals), averaged over a uniformly
    placed fix.\par
    Right: walks, of five, that named the fix; median: launches per walk. A launch is one natural trial.
    Fixed retry's $M$ by row: 20, 20, 150, 150, 10{,}921 and 5 (Appendix~\ref{sec:appendix-baseline}).\par
    \smallskip
    \chNote{a}{Case A's fix sits next to the fixed end of both its ranges, so most queried commits are vulnerable and
      hit quickly.}
    \chNote{b}{Capped fixed retry named commits 0--19 of 67.}
    \chNote{c}{Bayesect stops at a launch cap, 800 capped and 40{,}000 uncapped, which every capped walk and four of
      five uncapped ones reached.}
    \chNote{d}{Two searches of 8 trials (Table~\ref{tab:bisect-summary}'s \texttt{crun} rows); the second found commit
      27, the fix the walks are scored against.}\par}
\end{table}

\begin{figure}[!t]
\centering
\pgfplotstableread[col sep=comma]{figures/data/bisect_runs95.csv}\chRuns
\begin{tikzpicture}[ch]
\begin{axis}[chronos plot,
  width=6.3cm, height=3.3cm,
  xmode=log, ymode=log,
  xmin=0.0042, xmax=0.7, ymin=3, ymax=9000,
  xtick={0.005,0.01,0.03,0.1,0.3,0.6},
  xticklabels={0.5,1,3,10,30,60},
  ytick={10,100,1000}, yticklabels={10,100,{1,000}},
  xlabel={hit rate $q$ (\%)},
  ylabel={runs per bisection\\to be right 95\%},
  ylabel style={align=center, yshift=-2pt},
  xlabel style={yshift=2pt},
  unbounded coords=jump,
  legend columns=2,
  legend cell align=left,
  legend style={at={(0.5,-0.36)}, anchor=north, draw=none, fill=none, font=\scriptsize, text=ChInk,
    column sep=6pt, row sep=-1pt, /tikz/every even column/.append style={column sep=10pt}},
]
  \addplot[ChInkSoft, line width=0.55pt, mark=diamond*, mark size=1.9pt, mark options={fill=white}]
    table[x=q, y=syzbot_n19] {\chRuns};
  \addlegendentry{syzbot's rule}
  \addplot[ChInkSoft, line width=0.55pt, dash pattern=on 2.4pt off 1.2pt, mark=triangle*, mark size=2.4pt,
    mark options={solid, fill=white, draw=ChInkSoft, line width=0.5pt}]
    table[x=q, y=v8_n19] {\chRuns};
  \addlegendentry{V8 flake bisect}
  \addplot[ChAttack, line width=0.6pt, mark=*, mark size=1.4pt, mark options={fill=ChAttack, draw=white, line width=0.25pt}]
    table[x=q, y=bayesect_n19] {\chRuns};
  \addlegendentry{Git Bayesect}
  \addplot[ChInkSoft, line width=0.55pt, mark=square*, mark size=1.2pt, mark options={fill=white}]
    table[x=q, y=fixedretry_n19] {\chRuns};
  \addlegendentry{fixed retry}
  \addplot[ChAttack, line width=0.6pt, mark=o, mark size=1.4pt, mark options={line width=0.45pt}]
    table[x=q, y=pbacapacity_n19] {\chRuns};
  \addlegendentry{Bayesian, given $q$}
  \addplot[ChInk, line width=0.7pt, dash pattern=on 3pt off 1.5pt] table[x=q, y=floor_n19] {\chRuns};
  \addlegendentry{bound, bisectors without hold}
  \addplot[ChLabel, line width=0.9pt] table[x=q, y=chronos_n19] {\chRuns};
  \addlegendentry{Chronos}
\end{axis}
\end{tikzpicture}
\caption{\textbf{Syzbot's~\cite{syzkaller2026bisect} and V8's~\cite{v8flakebisect} rules need several times the runs of the dashed bound
  (Proposition~\ref{prop:bisect-bound}). Chronos needs about six trials at any rate.}
  Means over 2{,}000 simulated bisections per point on 19 commits, with the fix equally likely at any of them, every
  vulnerable commit hitting at rate $q$, and each method at its cheapest setting. Syzbot's rule reaches 95\% only from
  $q = 25\%$.}
\Description{A log-log plot of runs per bisection needed to be right 95 percent of the time against the hit rate q,
  from 0.5 to 60 percent, on 19 commits. A dashed bound falls from about 1,400 runs at 0.5 percent to about 9 at 60
  percent. Bayesian bisection given q runs just above it, fixed retry about 1.4 times above, Git Bayesect and V8's flake
  bisect about three times above, and syzbot's rule appears only from 25 percent, at about 85 to 360 runs. A
  flat magenta line at about 6 trials marks Chronos at every rate.}
\label{fig:bisect-runs95}
\end{figure}
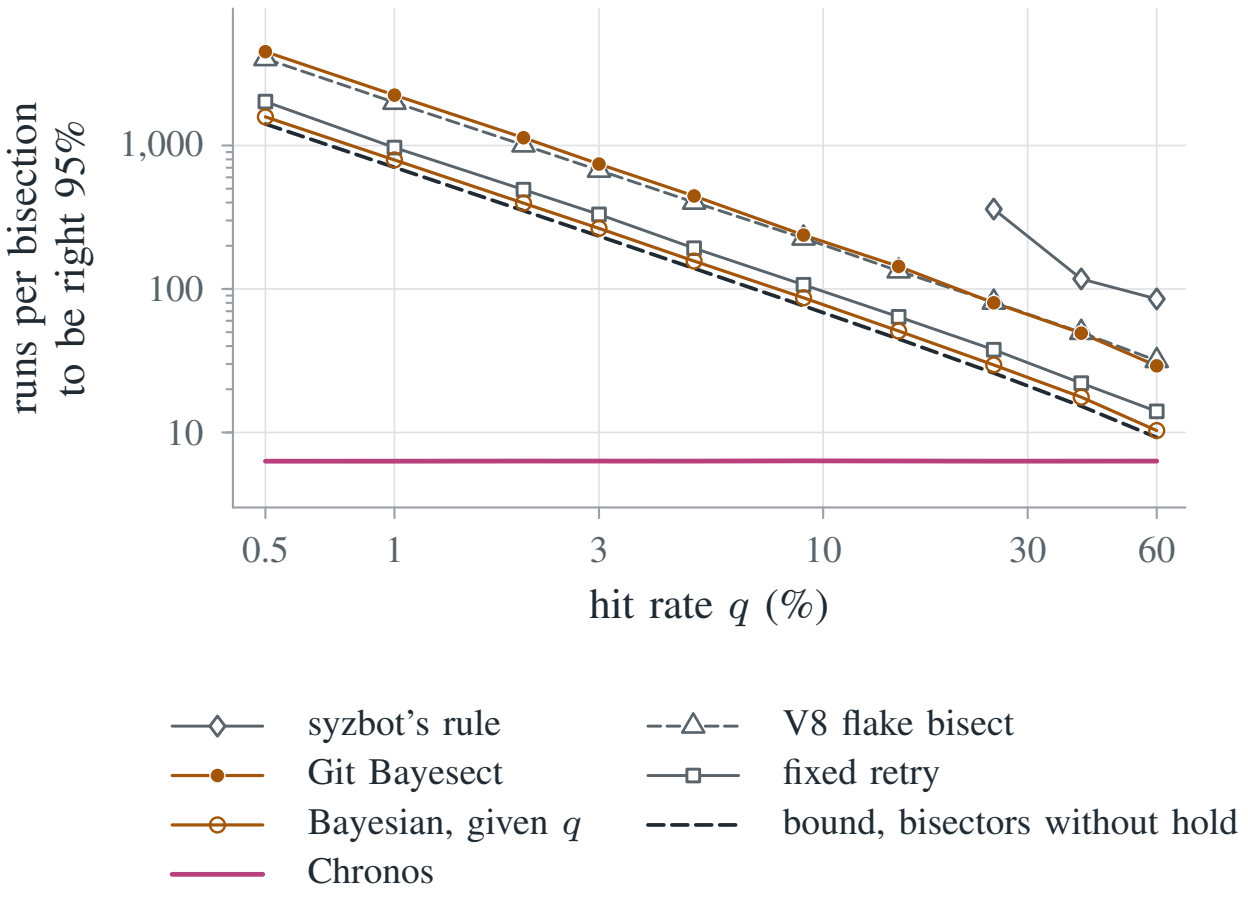

\paragraph{Retry budgets}
In the fixed-retry campaign of Table~\ref{tab:natural-rates}, one trial per commit named the fix in 0 of 10
  bisections, and $M{=}50$ in 10 of 10.
Rarer races cost far more:
  at a 1\% hit rate, even Bayesian bisection that knows the rate needs about 790 runs on average for 95\% accuracy
  on 19 commits (Fig.~\ref{fig:bisect-runs95}).

Replayed on our recorded natural runs, syzbot's rule named the fix in 48--62\% of bisections.
The cheapest setting of any method that named it in 95\% of bisections was fixed retry with $M$ of 19 to 27, at
  33--42 runs.

\paragraph{Across races}
Table~\ref{tab:bisect-races} (Appendix~\ref{sec:appendix-baseline}) gives Proposition~\ref{prop:bisect-bound}'s bound for the five searches at their original
  campaigns' natural rates (unindented rows): 5 to 99 times the trials Chronos needed (\texttt{crun}: at least 1{,}584 natural trials against 16), 12 times in
  geometric mean, although Chronos's counts
  include its endpoint checks and the bound counts only the search between them.
In those campaigns, natural trials at vulnerable revisions succeeded at rates of 0.65--20.4\%; none succeeded
  at fixed revisions.
The natural rate is a property of the attack's engineering and of the host, not of the race: with 19 victims and four
  swappers per launch, \texttt{crun}'s rate rose from 0.65\% to as much as 29\%, and with one victim it ranged from 0.10\% to 1.6\% on identical VMs, while Chronos's 16 trials did not change.
\texttt{crun} refused in 1{,}007 of 2{,}000 natural launches at 0.11, and in 10 of 50 at 0.12, there with the fix's
  own message.
\texttt{youki} and case D refused at their fixed endpoints too: 24/50 and 18/50, against 0/200 and 60/200 at their
  vulnerable endpoints.
These refusals show that the hit/miss bound does not cover every observable natural-run outcome.
Case C needs no bound, since a final-state check separates its versions (\S~\ref{sec:eval-proxy}).

\paragraph{Hosted CI}
\label{sec:eval-hosted-ci}
On each of 30 standard GitHub-hosted Ubuntu~24.04 runners~\cite{githubHostedRunners}, Chronos tested the vulnerable
  and fixed releases for CVE-2021-30465 (runc rc94 and rc95) and CVE-2025-31133 (v1.2.7 and v1.2.8) once, with the
  same Race Specifications as in reproduction and bisection.
All 120 labels matched the expected ones, none was indeterminate, and replaying each report gave the same label.
Excluding the build, the median run took 0.12--0.40\,s per endpoint, and the slowest endpoint's 95th percentile was
  0.75\,s, on each runner's stock Linux~6.17 kernel, without a nested VM or a custom kernel.
On Ubuntu~22.04 and 26.04 (April 2026) runners, with Linux~6.8 and 7.0, 8/8 labels were correct.

\subsection{RQ4: How Do Race Fixes Fail?}
\label{sec:eval-revealed}

Each finding in Table~\ref{tab:findings} needs a label for every release, which a natural run cannot give: at
  \texttt{crun}'s natural rate, measured as low as 0.10\%, a lost race looks like a fix.
We ran the specifications of cases A, C and D and of CVE-2024-45310 against the \texttt{runc} releases since
  \texttt{v1.0.0-rc1} and the \texttt{crun} and \texttt{youki} releases of Table~\ref{tab:release-sweep}.
Fig.~\ref{fig:fix-lifecycle} records which of its fixes we ran against their own attacks and against distribution
  packages.
We compared each label with the advisory's affected releases and the tracker's status.
Because vendors often block only the path that a proof of concept shows~\cite{stone2023zerodays}, we read each fix
  for paths it
  still resolves by name and held the attack, or a variant of it, at each one (Fig.~\ref{fig:fix-lifecycle}).
A stronger attack did not break case A's fix: when the helper exchanged the parent of the checked directory,
  rc94 was redirected in 20 of 20 runs and rc95's mount still covered the checked directory in 20 of 20.
We reported the new findings to their maintainers
  (Appendix~\ref{sec:appendix-ethics}).

\providecommand{\flX}{\tikz[baseline=-0.6ex]{\draw[ChAttack, line width=0.9pt] (-1.8pt,-1.8pt)--(1.8pt,1.8pt) (-1.8pt,1.8pt)--(1.8pt,-1.8pt);}}
\providecommand{\flNT}{\textcolor{ChGuide}{--}}
\providecommand{\flNA}{\textcolor{ChGuide}{n/a}}
\providecommand{\flUnrel}{\textcolor{ChGuide}{unrel.}}
\providecommand{\flH}[1]{\rotatebox{90}{#1}}
\providecommand{\flGroup}[1]{\multicolumn{6}{@{}l}{\textit{#1}}\\}
\begin{figure}[t]
  \centering
  \scriptsize
  \setlength{\tabcolsep}{2pt}
  \begin{tabular*}{\columnwidth}{@{\extracolsep{\fill}}ll cccc@{}}
    \toprule
    \textbf{Fix} & \textbf{Release} & \flH{Own attack} & \flH{Variants} & \flH{Advisory} & \flH{Packages} \\
    \midrule
    \flGroup{runc}
    \textbf{A} CVE-2021-30465 & rc95 & \chyes & \chyes & \chyes & \chyes\,7/7 \\  %
    \textbf{C} CVE-2023-27561 & 1.1.5 & \chyes & \chyes & \chyes & \chyes\,7/7 \\  %
    \textbf{D} CVE-2025-31133 & 1.2.8 & \chyes & \chyes & \chyes & \chpart\,6/7 \\  %
    CVE-2024-45310 & 1.1.14 & \chyes & \chyes & \chyes & \chpart\,6/7 \\  %
    CVE-2019-19921 & rc10 & \chyes & \chyes & \chpart$^{a}$ & \flNT \\  %
    CVE-2025-52565 & 1.2.8 & \chyes & \flNT & \chpart & \chpart\,6/7 \\  %
    \flGroup{Other container tools}
    \textbf{B} NVIDIA hook & 1.17.4 & \chyes & \chyes$^{b}$ & \chyes & \flNT \\  %
    crun, A's attack & 0.12 & \chyes & \chyes & \flNA$^{c}$ & \chyes\,5/5 \\  %
    youki, A's attack & v0.5.7 & \chyes & \flNT & \flNA$^{c}$ & \flNA \\  %
    kubelet & v1.21.5 & \chyes & \chyes & \chyes & \flNA \\  %
    Docker & 18.09.7 & \chyes & \flNT & \chpart & \flNT \\  %
    \flGroup{System software}
    shadow & 4.12.2 & \chyes & \chyes & \chpart & \chpart\,4/7 \\  %
    rpm & 4.18.0 & \chyes & \flX & \chpart & \chpart\,5/7 \\  %
    rpm, Debian 12 & deb12u1 & \flNA & \flNA & \flX & \flX \\  %
    chrony & 4.8 & \flNT & \flX & \chpart & \flNT \\  %
    chrony, proposed fix~\cite{chronySocketFix2026} & patch & \chyes & \chyes & \flUnrel & \flUnrel \\  %
    Cyrus IMAP & 3.12.1 & \chyes & \chyes & \chpart & \chpart\,1/6 \\  %
    Apache httpd & 2.4.10 & \chyes & \chyes & \chyes & \chyes\,7/7 \\  %
    rsync & 3.4.3 & \chyes & \flNT & \chyes & \chyes\,7/7 \\  %
    apport & 2.33.0 & \chyes & \flNT & \chyes$^{d}$ & \chyes\,2/2 \\  %
    QEMU & 7.0.0 & \chyes & \flNT & \chpart & \flNT \\  %
    sudo & 1.9.5 & \chyes & \flNT & \chyes$^{d}$ & \chyes\,7/7 \\  %
    \flGroup{Linux kernel}
    CVE-2022-49919 (\texttt{nf\_tables}) & fix & \chyes & \flNT & \chyes & \flNT \\  %
    iommufd \texttt{6f9c4d8c} & fix & \chyes & \flNT & \chpart & \flNT \\  %
    \midrule
    \multicolumn{2}{@{}l}{Passed, of tested} & 22/22 & 12/14 & 11/21 & 7/14 \\
    \bottomrule
  \end{tabular*}

  \vspace{3pt}
  {\scriptsize\raggedright \chyes{} passed; \chpart{} partly; \flX{} failed; \flNT{} not tested; \flNA{} does not apply; \flUnrel{} unreleased.\par
    \smallskip
    Each row is a fix as its vendor shipped it: an upstream release or patch, or a distribution's package. Own attack: the fix's own proof of concept. Variants: the attack moved to paths the fix still resolves by name. Advisory: the advisory or tracker lists every exposed release. Packages: K of the N supported releases tested are labeled fixed by held runs of their packages (November 2026); the supported releases are Debian~12, Ubuntu~22.04 and 24.04, Rocky~8 and 9, Amazon Linux~2023 and openSUSE Leap~16.0. Each cell shows the record as it stood before any report of ours.\par
    \smallskip
    \chNote{a}{The releases that re-exposed CVE-2019-19921's race were listed 21 months later, under CVE-2023-27561 (Table~\ref{tab:release-sweep}).}
    \chNote{b}{The variant needs the opt-in compat mode that restores the mount (RQ3), which only the host's administrator can set.}
    \chNote{c}{runc's CVE records cover runc only.}
    \chNote{d}{The stated range also lists releases that carry the fix: sudo's 1.8.32 and apport's 2.28.2 to 2.28.4.}
    \par}
  \caption{\textbf{Of 24 fixes, 3 failed a stage: 2 in their code, 1 only in its records or packages.} Read across a row for one fix; Table~\ref{tab:findings} says what each cross found.}
  \label{fig:fix-lifecycle}
\end{figure}

\paragraph{Release sweep}
Reused unchanged, the specifications of cases A, C and D labeled every \texttt{runc} release since
  \texttt{v1.0.0-rc1} definitively.
With only their paths changed, and case C's rule naming each runtime's refusal, the specifications of cases A and C
  and of CVE-2024-45310 labeled the \texttt{crun} and \texttt{youki} releases too; case D's change cannot be made on
  81 \texttt{crun} releases, which mask with the host's \texttt{/dev/null} before \texttt{pivot\_root}, and on \texttt{youki} it labeled
  vulnerable exactly the 20 releases the advisory lists.
Control targets for cases A and D, which start but exit without making the held call, were labeled indeterminate.
All 371 \texttt{runc} release--specification labels, those of cases A, C and D and of CVE-2024-45310, matched
  advisory status (Table~\ref{tab:release-sweep}).
Repeating the sweep reproduced every label.

\begin{table}[t]
  \caption{\textbf{Every \texttt{runc} label, and every definitive \texttt{crun} and \texttt{youki} label that an
    advisory covers, matched its advisory; \texttt{crun}'s and \texttt{youki}'s earlier releases were also exposed to
    the attacks of two \texttt{runc} CVEs.}}
  \label{tab:release-sweep}
  \centering
  \scriptsize
  \setlength{\tabcolsep}{3pt}
  \begin{tabular*}{\columnwidth}{@{\extracolsep{\fill}}llrrrrr@{}}
    \toprule
    & & & \multicolumn{3}{c}{\textbf{Labels}} & \\
    \cmidrule(lr){4-6}
    \textbf{Runtime} & \textbf{Spec} & \textbf{Releases} & \textbf{Vuln.} & \textbf{Fixed} & \textbf{Indet.} &
      \shortstack[r]{\textbf{Advisory}\\\textbf{matches}} \\
    \midrule
    \texttt{runc}  & A, C, D & 75 & 92 & 133 & 0 & 225/225$^{a}$ \\
                   & 45310 & 73 & 74 & 72 & 0 & 146/146 \\
    \addlinespace
    \texttt{crun}  & A & 96 & 17 & 79 & 0 & -- \\
                   & C & 96 & 8 & 79 & 9$^{b}$ & 87/87$^{c}$ \\
                   & 45310 & 96 & 34 & 157 & 1$^{d}$ & -- \\
                   & D & 96 & 0 & 2 & 94$^{e}$ & -- \\
    \addlinespace
    \texttt{youki} & A & 23 & 20 & 3 & 0 & -- \\
                   & C & 23 & 18 & 5 & 0 & 23/23$^{f}$ \\
                   & 45310 & 23 & 40 & 6 & 0 & -- \\
                   & D & 23 & 20 & 3 & 0 & 23/23$^{g}$ \\
    \bottomrule
  \end{tabular*}

  \vspace{3pt}
  {\scriptsize\raggedright --: no advisory names the attack.\par
    \smallskip
    The sweep covers every tag of \texttt{runc} (73), \texttt{crun} and \texttt{youki} through October 2026, and the
    \texttt{runc} builds of Ubuntu 20.04 and 24.04. The 45310 rows are CVE-2024-45310's directory and file
    specifications (Table~\ref{tab:cases}), two labels per release. Advisory matches: definitive labels that agree
    with the advisories.\par
    \smallskip
    \chNote{a}{Case C has two \texttt{runc} advisories: CVE-2019-19921 through rc9 and CVE-2023-27561 from rc95 to
      1.1.4.}
    \chNote{b}{\texttt{crun}'s C at v0.2 exits early, and at 0.10 to 0.11 mounts procfs at a misjoined path outside
      the root.}
    \chNote{c}{\texttt{crun}'s CVE-2019-18837 lists every release before 0.10.5 for case C's attack.}
    \chNote{d}{\texttt{crun}'s 45310 at 0.14 makes no held call.}
    \chNote{e}{Case D's change cannot be made on \texttt{crun} (81 releases), \texttt{crun} makes no held call (11),
      or the held call restarts after the window (2).}
    \chNote{f}{\texttt{youki}'s CVE-2025-54867 lists every release before v0.5.5 for case C's attack.}
    \chNote{g}{\texttt{youki}'s CVE-2025-62161 lists every release before v0.5.7 for case D's attack.}\par}
\end{table}

\begin{table*}[t]
  \caption{\textbf{Three fixes or packages that their records called fixed or mitigated were labeled vulnerable; the labels also
    caught runc's known reintroduction of an older CVE.} Records are
    quoted as they stood before any report of ours.}
  \label{tab:findings}
  \centering
  \scriptsize
  \setlength{\tabcolsep}{3pt}
  \begin{tabularx}{\textwidth}{@{}>{\raggedright\arraybackslash}p{0.24\textwidth}>{\raggedright\arraybackslash}X%
    >{\raggedright\arraybackslash}p{0.27\textwidth}@{}}
    \toprule
    \textbf{Record compared} & \textbf{What the labels showed} & \textbf{Evidence} \\
    \midrule
    \tabgroup{3}{A record called the build fixed or mitigated}
    chrony 4.8's release notes~\cite{chronyNews2025}: ``Hide chronyc socket to mitigate unsafe permissions change''
      (August 2025)
      & Chrony 4.8 and 4.9 retain a client-socket race that redirected root's permission change in the tested
      unconfined configuration. Seccomp and SELinux confinement can block this path~\cite{chronySocketReport2026}.
      & Held: 4.8 and 4.9 vulnerable 10/10 each; 4.9 with the maintainer's proposed fix~\cite{chronySocketFix2026}
      fixed 10/10. Unheld, the attack won once in 1{,}200 runs. \\
    \addlinespace
    NVD: ``Fixed in rpm v4.18.0'' (CVE-2021-35938; read November 2026)
      & rpm 4.18.0's fix missed the path that changes a file's mode when an upgrade leaves its content unchanged.
      & Held: 4.18.0 vulnerable as 4.17.1, 4.18.1 fixed. \\
    \addlinespace
    Debian's tracker: ``bookworm 4.18.0+dfsg-1+deb12u1 fixed'' (April 2026; amended in November 2026)
      & Debian~12's tested package, \texttt{4.18.0+dfsg-1+deb12u1}, retains that path~\cite{debianRpmTracker2026}.
      & Held: 3/3 vulnerable (\S~\ref{sec:limitations}). \\
    \tabgroup{3}{A known regression}
    CVE-2019-19921's advisory: ``fixed in \texttt{1.0.0-rc10}''; CVE-2023-27561 (March 2023): rc95 to 1.1.4
      & \texttt{runc}'s \texttt{0ca91f44} (rc95) repaired CVE-2021-30465 and reintroduced CVE-2019-19921
      (Fig.~\ref{fig:one-commit-two-labels}); the exposure was listed only 21 months after rc95, as CVE-2023-27561.
      & Replay labels \texttt{0ca91f44} vulnerable for case C's contract and fixed for case A's, and C's
      introduction search names it (Table~\ref{tab:bisect-summary}). \\
    \bottomrule
  \end{tabularx}
\end{table*}

\paragraph{Distribution packages}
The specifications that labeled \texttt{crun}'s releases ran unchanged against the packaged \texttt{crun} binaries of
  18 distribution releases (Debian, Ubuntu, Fedora, Rocky Linux, Alpine and openSUSE), three held runs each.
Every package received the label of the upstream build of its version: fixed for cases A and C and CVE-2024-45310, and
  for case D fixed on the packages of \texttt{crun} 1.30 and later and indeterminate on the rest.
Debian~12's \texttt{rpm} package, by contrast, was labeled vulnerable while Debian's tracker listed it as fixed
  (Table~\ref{tab:findings}).

\paragraph{rpm: a fix that needed positive evidence}
CVE-2021-35938 lets the account that owns an installation directory redirect rpm's metadata changes to another
  file~\cite{rpmCve202135938}.
Release 4.18.0 shipped a descriptor-based fix~\cite{rpm4180Notes}, but on its \texttt{FA\_TOUCH} path, used when an
  upgrade leaves a file's content unchanged, it still calls \texttt{fchmodat} on the file's name.
Under the held swap, 4.18.0 changed the canary's mode from 0600 to 0666, as 4.17.1 did.
Release 4.18.1 calls \texttt{fchmod} on the descriptor it opened, so the original file changed and the canary kept
  its mode: positive evidence for the fixed label.
Without the hold, 23 of 100 natural runs on 4.17.1 and 17 of 100 on 4.18.0 reached the effect.
A natural run that loses the race shows neither and cannot tell 4.18.0 from 4.18.1.

\paragraph{Where the fixes failed}
Fig.~\ref{fig:fix-lifecycle} counts full passes among the fixes tested at each stage: 22 of 22 stopped their own
  attack and 12 of 14 stopped the variants.
For the records and packages, 11 of 21 had an advisory or tracker that listed every exposed release, and 7 of 14
  were labeled fixed in every package we ran.
In our corpus, the three issues are two incomplete fixes and a package that kept one: the fixes in chrony and
  rpm, and Debian~12's rpm package, which kept rpm's.
In all three, a record listed the build as fixed or, for chrony, as mitigated, and the build still resolves a path by
  name.
A later commit closes that path: \texttt{8b5cce9a} in rpm~4.18.1, which Debian~12's package lacks, and for chrony a
  proposed commit that no release carries~\cite{chronySocketFix2026}.
The later commits show where each path was closed, and the labels show which builds are vulnerable.
Where a runtime closed the race in two commits, as \texttt{crun} did, one search per held call named each
  (\S~\ref{sec:eval-bisect}).
PATROL's study of container runtimes traces path-misresolution flaws that recurred after earlier
  patches~\cite{patrol2023}.

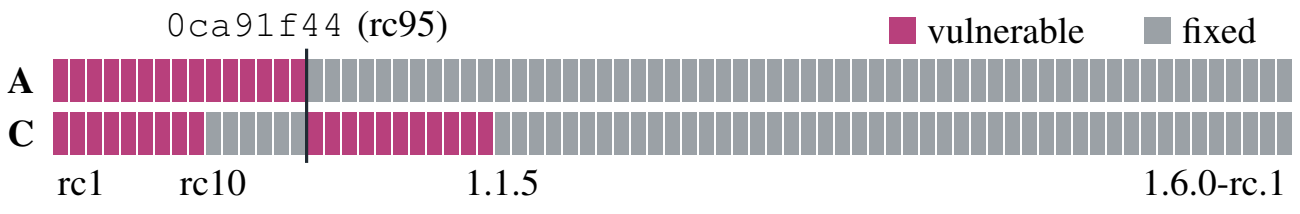
\begin{figure}[t]
  \centering
  \begin{tikzpicture}[x=1pt, y=1pt, font=\scriptsize]
    \fill[ChLabel] (0.25,10.00) rectangle (2.95,18.00);  %
    \fill[ChLabel] (3.45,10.00) rectangle (6.15,18.00);  %
    \fill[ChLabel] (6.65,10.00) rectangle (9.35,18.00);  %
    \fill[ChLabel] (9.85,10.00) rectangle (12.55,18.00);  %
    \fill[ChLabel] (13.05,10.00) rectangle (15.75,18.00);  %
    \fill[ChLabel] (16.25,10.00) rectangle (18.95,18.00);  %
    \fill[ChLabel] (19.45,10.00) rectangle (22.15,18.00);  %
    \fill[ChLabel] (22.65,10.00) rectangle (25.35,18.00);  %
    \fill[ChLabel] (25.85,10.00) rectangle (28.55,18.00);  %
    \fill[ChLabel] (29.05,10.00) rectangle (31.75,18.00);  %
    \fill[ChLabel] (32.25,10.00) rectangle (34.95,18.00);  %
    \fill[ChLabel] (35.45,10.00) rectangle (38.15,18.00);  %
    \fill[ChLabel] (38.65,10.00) rectangle (41.35,18.00);  %
    \fill[ChLabel] (41.85,10.00) rectangle (44.55,18.00);  %
    \fill[ChLabel] (45.05,10.00) rectangle (47.75,18.00);  %
    \fill[ChGuide] (48.25,10.00) rectangle (50.95,18.00);  %
    \fill[ChGuide] (51.45,10.00) rectangle (54.15,18.00);  %
    \fill[ChGuide] (54.65,10.00) rectangle (57.35,18.00);  %
    \fill[ChGuide] (57.85,10.00) rectangle (60.55,18.00);  %
    \fill[ChGuide] (61.05,10.00) rectangle (63.75,18.00);  %
    \fill[ChGuide] (64.25,10.00) rectangle (66.95,18.00);  %
    \fill[ChGuide] (67.45,10.00) rectangle (70.15,18.00);  %
    \fill[ChGuide] (70.65,10.00) rectangle (73.35,18.00);  %
    \fill[ChGuide] (73.85,10.00) rectangle (76.55,18.00);  %
    \fill[ChGuide] (77.05,10.00) rectangle (79.75,18.00);  %
    \fill[ChGuide] (80.25,10.00) rectangle (82.95,18.00);  %
    \fill[ChGuide] (83.45,10.00) rectangle (86.15,18.00);  %
    \fill[ChGuide] (86.65,10.00) rectangle (89.35,18.00);  %
    \fill[ChGuide] (89.85,10.00) rectangle (92.55,18.00);  %
    \fill[ChGuide] (93.05,10.00) rectangle (95.75,18.00);  %
    \fill[ChGuide] (96.25,10.00) rectangle (98.95,18.00);  %
    \fill[ChGuide] (99.45,10.00) rectangle (102.15,18.00);  %
    \fill[ChGuide] (102.65,10.00) rectangle (105.35,18.00);  %
    \fill[ChGuide] (105.85,10.00) rectangle (108.55,18.00);  %
    \fill[ChGuide] (109.05,10.00) rectangle (111.75,18.00);  %
    \fill[ChGuide] (112.25,10.00) rectangle (114.95,18.00);  %
    \fill[ChGuide] (115.45,10.00) rectangle (118.15,18.00);  %
    \fill[ChGuide] (118.65,10.00) rectangle (121.35,18.00);  %
    \fill[ChGuide] (121.85,10.00) rectangle (124.55,18.00);  %
    \fill[ChGuide] (125.05,10.00) rectangle (127.75,18.00);  %
    \fill[ChGuide] (128.25,10.00) rectangle (130.95,18.00);  %
    \fill[ChGuide] (131.45,10.00) rectangle (134.15,18.00);  %
    \fill[ChGuide] (134.65,10.00) rectangle (137.35,18.00);  %
    \fill[ChGuide] (137.85,10.00) rectangle (140.55,18.00);  %
    \fill[ChGuide] (141.05,10.00) rectangle (143.75,18.00);  %
    \fill[ChGuide] (144.25,10.00) rectangle (146.95,18.00);  %
    \fill[ChGuide] (147.45,10.00) rectangle (150.15,18.00);  %
    \fill[ChGuide] (150.65,10.00) rectangle (153.35,18.00);  %
    \fill[ChGuide] (153.85,10.00) rectangle (156.55,18.00);  %
    \fill[ChGuide] (157.05,10.00) rectangle (159.75,18.00);  %
    \fill[ChGuide] (160.25,10.00) rectangle (162.95,18.00);  %
    \fill[ChGuide] (163.45,10.00) rectangle (166.15,18.00);  %
    \fill[ChGuide] (166.65,10.00) rectangle (169.35,18.00);  %
    \fill[ChGuide] (169.85,10.00) rectangle (172.55,18.00);  %
    \fill[ChGuide] (173.05,10.00) rectangle (175.75,18.00);  %
    \fill[ChGuide] (176.25,10.00) rectangle (178.95,18.00);  %
    \fill[ChGuide] (179.45,10.00) rectangle (182.15,18.00);  %
    \fill[ChGuide] (182.65,10.00) rectangle (185.35,18.00);  %
    \fill[ChGuide] (185.85,10.00) rectangle (188.55,18.00);  %
    \fill[ChGuide] (189.05,10.00) rectangle (191.75,18.00);  %
    \fill[ChGuide] (192.25,10.00) rectangle (194.95,18.00);  %
    \fill[ChGuide] (195.45,10.00) rectangle (198.15,18.00);  %
    \fill[ChGuide] (198.65,10.00) rectangle (201.35,18.00);  %
    \fill[ChGuide] (201.85,10.00) rectangle (204.55,18.00);  %
    \fill[ChGuide] (205.05,10.00) rectangle (207.75,18.00);  %
    \fill[ChGuide] (208.25,10.00) rectangle (210.95,18.00);  %
    \fill[ChGuide] (211.45,10.00) rectangle (214.15,18.00);  %
    \fill[ChGuide] (214.65,10.00) rectangle (217.35,18.00);  %
    \fill[ChGuide] (217.85,10.00) rectangle (220.55,18.00);  %
    \fill[ChGuide] (221.05,10.00) rectangle (223.75,18.00);  %
    \fill[ChGuide] (224.25,10.00) rectangle (226.95,18.00);  %
    \fill[ChGuide] (227.45,10.00) rectangle (230.15,18.00);  %
    \fill[ChGuide] (230.65,10.00) rectangle (233.35,18.00);  %
    \node[anchor=east, inner sep=1pt] at (-2,14.00) {\textbf{A}};
    \fill[ChLabel] (0.25,0.00) rectangle (2.95,8.00);  %
    \fill[ChLabel] (3.45,0.00) rectangle (6.15,8.00);  %
    \fill[ChLabel] (6.65,0.00) rectangle (9.35,8.00);  %
    \fill[ChLabel] (9.85,0.00) rectangle (12.55,8.00);  %
    \fill[ChLabel] (13.05,0.00) rectangle (15.75,8.00);  %
    \fill[ChLabel] (16.25,0.00) rectangle (18.95,8.00);  %
    \fill[ChLabel] (19.45,0.00) rectangle (22.15,8.00);  %
    \fill[ChLabel] (22.65,0.00) rectangle (25.35,8.00);  %
    \fill[ChLabel] (25.85,0.00) rectangle (28.55,8.00);  %
    \fill[ChGuide] (29.05,0.00) rectangle (31.75,8.00);  %
    \fill[ChGuide] (32.25,0.00) rectangle (34.95,8.00);  %
    \fill[ChGuide] (35.45,0.00) rectangle (38.15,8.00);  %
    \fill[ChGuide] (38.65,0.00) rectangle (41.35,8.00);  %
    \fill[ChGuide] (41.85,0.00) rectangle (44.55,8.00);  %
    \fill[ChGuide] (45.05,0.00) rectangle (47.75,8.00);  %
    \fill[ChLabel] (48.25,0.00) rectangle (50.95,8.00);  %
    \fill[ChLabel] (51.45,0.00) rectangle (54.15,8.00);  %
    \fill[ChLabel] (54.65,0.00) rectangle (57.35,8.00);  %
    \fill[ChLabel] (57.85,0.00) rectangle (60.55,8.00);  %
    \fill[ChLabel] (61.05,0.00) rectangle (63.75,8.00);  %
    \fill[ChLabel] (64.25,0.00) rectangle (66.95,8.00);  %
    \fill[ChLabel] (67.45,0.00) rectangle (70.15,8.00);  %
    \fill[ChLabel] (70.65,0.00) rectangle (73.35,8.00);  %
    \fill[ChLabel] (73.85,0.00) rectangle (76.55,8.00);  %
    \fill[ChLabel] (77.05,0.00) rectangle (79.75,8.00);  %
    \fill[ChLabel] (80.25,0.00) rectangle (82.95,8.00);  %
    \fill[ChGuide] (83.45,0.00) rectangle (86.15,8.00);  %
    \fill[ChGuide] (86.65,0.00) rectangle (89.35,8.00);  %
    \fill[ChGuide] (89.85,0.00) rectangle (92.55,8.00);  %
    \fill[ChGuide] (93.05,0.00) rectangle (95.75,8.00);  %
    \fill[ChGuide] (96.25,0.00) rectangle (98.95,8.00);  %
    \fill[ChGuide] (99.45,0.00) rectangle (102.15,8.00);  %
    \fill[ChGuide] (102.65,0.00) rectangle (105.35,8.00);  %
    \fill[ChGuide] (105.85,0.00) rectangle (108.55,8.00);  %
    \fill[ChGuide] (109.05,0.00) rectangle (111.75,8.00);  %
    \fill[ChGuide] (112.25,0.00) rectangle (114.95,8.00);  %
    \fill[ChGuide] (115.45,0.00) rectangle (118.15,8.00);  %
    \fill[ChGuide] (118.65,0.00) rectangle (121.35,8.00);  %
    \fill[ChGuide] (121.85,0.00) rectangle (124.55,8.00);  %
    \fill[ChGuide] (125.05,0.00) rectangle (127.75,8.00);  %
    \fill[ChGuide] (128.25,0.00) rectangle (130.95,8.00);  %
    \fill[ChGuide] (131.45,0.00) rectangle (134.15,8.00);  %
    \fill[ChGuide] (134.65,0.00) rectangle (137.35,8.00);  %
    \fill[ChGuide] (137.85,0.00) rectangle (140.55,8.00);  %
    \fill[ChGuide] (141.05,0.00) rectangle (143.75,8.00);  %
    \fill[ChGuide] (144.25,0.00) rectangle (146.95,8.00);  %
    \fill[ChGuide] (147.45,0.00) rectangle (150.15,8.00);  %
    \fill[ChGuide] (150.65,0.00) rectangle (153.35,8.00);  %
    \fill[ChGuide] (153.85,0.00) rectangle (156.55,8.00);  %
    \fill[ChGuide] (157.05,0.00) rectangle (159.75,8.00);  %
    \fill[ChGuide] (160.25,0.00) rectangle (162.95,8.00);  %
    \fill[ChGuide] (163.45,0.00) rectangle (166.15,8.00);  %
    \fill[ChGuide] (166.65,0.00) rectangle (169.35,8.00);  %
    \fill[ChGuide] (169.85,0.00) rectangle (172.55,8.00);  %
    \fill[ChGuide] (173.05,0.00) rectangle (175.75,8.00);  %
    \fill[ChGuide] (176.25,0.00) rectangle (178.95,8.00);  %
    \fill[ChGuide] (179.45,0.00) rectangle (182.15,8.00);  %
    \fill[ChGuide] (182.65,0.00) rectangle (185.35,8.00);  %
    \fill[ChGuide] (185.85,0.00) rectangle (188.55,8.00);  %
    \fill[ChGuide] (189.05,0.00) rectangle (191.75,8.00);  %
    \fill[ChGuide] (192.25,0.00) rectangle (194.95,8.00);  %
    \fill[ChGuide] (195.45,0.00) rectangle (198.15,8.00);  %
    \fill[ChGuide] (198.65,0.00) rectangle (201.35,8.00);  %
    \fill[ChGuide] (201.85,0.00) rectangle (204.55,8.00);  %
    \fill[ChGuide] (205.05,0.00) rectangle (207.75,8.00);  %
    \fill[ChGuide] (208.25,0.00) rectangle (210.95,8.00);  %
    \fill[ChGuide] (211.45,0.00) rectangle (214.15,8.00);  %
    \fill[ChGuide] (214.65,0.00) rectangle (217.35,8.00);  %
    \fill[ChGuide] (217.85,0.00) rectangle (220.55,8.00);  %
    \fill[ChGuide] (221.05,0.00) rectangle (223.75,8.00);  %
    \fill[ChGuide] (224.25,0.00) rectangle (226.95,8.00);  %
    \fill[ChGuide] (227.45,0.00) rectangle (230.15,8.00);  %
    \fill[ChGuide] (230.65,0.00) rectangle (233.35,8.00);  %
    \node[anchor=east, inner sep=1pt] at (-2,4.00) {\textbf{C}};
    \draw[ChInk, line width=0.6pt] (48.00,-1.5) -- (48.00,20.00);
    \node[anchor=south, inner sep=1pt] at (48.00,20.00) {\texttt{0ca91f44} (rc95)};
    \node[anchor=north west, inner sep=1pt] at (0,-1.5) {rc1};
    \node[anchor=north, inner sep=1pt] at (30.40,-1.5) {rc10};
    \node[anchor=north, inner sep=1pt] at (84.80,-1.5) {1.1.5};
    \node[anchor=north east, inner sep=1pt] at (233.60,-1.5) {1.6.0-rc.1};
    \fill[ChLabel] (157.60,21.00) rectangle ++(5,5);
    \node[anchor=west, inner sep=1pt] at (163.60,23.50) {vulnerable};
    \fill[ChGuide] (205.60,21.00) rectangle ++(5,5);
    \node[anchor=west, inner sep=1pt] at (211.60,23.50) {fixed};
  \end{tikzpicture}
  \caption{\textbf{\texttt{runc}'s \texttt{0ca91f44} is fixed for case A (CVE-2021-30465) and vulnerable for case C (CVE-2023-27561).}
    One cell per release tag.
    C was vulnerable until rc10, and again from rc95 until 1.1.5, 22 months later.}
  \label{fig:one-commit-two-labels}
\end{figure}

\paragraph{Regression coverage}
A regression suite that keeps case C's Race Specification fails on \texttt{runc}'s \texttt{0ca91f44}, the commit that
  undid CVE-2019-19921's fix (Fig.~\ref{fig:one-commit-two-labels}; Table~\ref{tab:findings}).

\section{Discussion}
\label{sec:discussion}

\noindent\textbf{Implications.}\label{sec:implications}
A natural run that misses a race is weak evidence about a fix (\S~\ref{sec:eval-revealed}): repeated misses only
  bound the race's rate, and Proposition~\ref{prop:bisect-bound} bounds what that costs a search.
A uniform finite-sample guarantee for statistical retesting~\cite{Henderson2023FlakeAware,GitBayesect} therefore
  requires a known positive lower bound on that rate; even then, the conclusion is probabilistic.
Where the decisive ordering is known, one held run gives a definitive or indeterminate label.
An advisory can list every commit of a race fix, and a project can keep the fix's Race Specification in CI, so that
  a later commit that undoes the fix fails the check.
A distribution can record which of those commits a backport holds, and call a package fixed for a race when that
  race's specification labels it fixed.
Debian~12's package carried rpm~4.18.0's fix and was labeled vulnerable (\S~\ref{sec:eval-revealed}).

\noindent\textbf{Limitations.}\label{sec:limitations}
The seccomp engine needs a seccomp-trappable use syscall and \codebreak{SECCOMP_USER_NOTIF_FLAG_CONTINUE}
  (Linux~5.5, or a backport~\cite{LinuxSeccompUnotify}), and the \texttt{mount} family needs \texttt{listmount} and
  \texttt{statmount} (Linux~6.8) to tell mounts apart.
The Linux~7.0 release sweep measures labels under one kernel feature set.
  Fig.~\ref{fig:fix-lifecycle}'s package runs used each release's own packages and userland on kernels from 5.14 to
  7.3 (the mount family's on 6.8 and later), often newer than the release ships.
  On an older kernel a target may take another path.
The supervisor fails a run whose target creates an \texttt{io\_uring} ring, which would bypass syscall-entry
  observation, but cannot see an inherited or received ring; the labels assume the target uses none.
A target that installs its own seccomp notification listener does not displace Chronos's: the kernel refuses
  the second listener, and on \texttt{runc} the labels were unchanged or the run was indeterminate.
A concurrent propagated mount from another process could mimic the held call's arrival at a candidate point, so the
  guarantee assumes no such mount.

A release that keeps the declared use safe but brings the bug back through another, undeclared use is labeled
  fixed, unless an outcome rule checks the effect, as case D's does.
By contrast, the reverts of \S~\ref{sec:eval-window} that moved the bug made no held call, so they ended
  indeterminate.
Where a fixed label rests on a call never made (case B), a use moved to a path the pattern does not match is labeled
  fixed, and the label covers the configuration the setup creates, such as NVIDIA's default.
A label also trusts its specification: the hold binds the first matching call (\S~\ref{sec:lifecycle}), the held
  object stands for the checked one unless the check call is declared (\S~\ref{sec:freeze-soundness}), and a
  refusal counts only with the fix's declared message.

Because a syscall-entry hold acts before the kernel resolves any name, a race between the steps of one pathname
  walk needs a hold inside the kernel, such as the kernel-instruction hold.
Some races cannot be held at all: suspending the use may also stop the competing action, for example when both
  need the same lock.

Each case starts from a known race, since Chronos labels only the race that its specification declares.
From there, its specification and variants of its attack labeled the releases, distribution packages and other
  runtimes of RQ4, where all of RQ4's issues were found.

\section{Related Work}
\label{sec:related}

Chronos holds a known race's use, mainly in unmodified userspace programs of container infrastructure, while an
  attacker makes the declared change, in most cases to a name the use will resolve, and labels each version by the
  selected use's effect or a declared outcome rule.
Most of the closest tools target other races, inside one program's threads, between cooperating processes or
  inside the kernel. Our contribution is a single-trial regression oracle under stated assumptions, together
  with its evaluation across repairs, cases and software histories.
Table~\ref{tab:related-positioning} summarizes capabilities demonstrated for these systems' different objectives.

\noindent\textbf{Name races and their defenses.}
STING~\cite{sting2012} comes closest: it attacks name resolutions system-wide through a modified kernel to find new
  flaws, and records each attacked call as vulnerable or defended.
A defended call shows that one attack failed, not that a version fixed a declared race, and STING searches no
  versions.
Process Firewalls~\cite{processFirewalls2013} and PATROL~\cite{patrol2023} block such attacks in kernel policy, so a
  denial measures the policy, not the program.
Bulkhead~\cite{bulkhead2026} combines static analysis and LLM reasoning to find container path vulnerabilities,
  and validates generated repairs through model checking and runtime tests.
LXD and Sysbox use seccomp notification to mediate container syscalls~\cite{lxdSyscallInterception,
  sysboxSyscallInterception}, as Janus held syscalls to enforce a sandbox~\cite{garfinkel2003traps}; none orders a
  race.
Earlier work mapped racy pairs of file-system calls~\cite{wei2005tocttou} and made path resolution resist
  them~\cite{tsafrir2008portably,chari2010where}.

\noindent\textbf{Other races.}
RacePro~\cite{racepro2011}, RRF~\cite{rrf2016} and SimEvo~\cite{simevo2017} reorder accesses between cooperating
  processes, through a kernel recorder, sleeps compiled into the source or a full-system simulator, and judge harm
  with application checkers. A passing controlled test provides evidence about the exercised behavior, within
  the scope of those checks.
CONCURRIT~\cite{concurrit2013} and CHESS~\cite{chess2008} perform instrumented thread-schedule testing;
  CONCURRIT also supports multi-process system tests.
A script fixed to one schedule can serve as a regression test, with assertion failures, crashes, deadlocks,
  livelocks or attached monitors as its oracle.
Chronos holds the declared call of an unmodified target at the OS boundary while a separate attacker process makes
  the change, and labels a security effect that need not crash, fixed only on positive evidence.
CARDSHARK~\cite{cardshark2024} (\chpt{1} in Fig.~\ref{fig:intercept-map}) and SysPro~\cite{SysPro2026} (\chpt{2})
  reproduce known races with aligned or delayed calls, so a miss still leaves a version unlabeled.
Kernel concurrency fuzzers explore internal interleavings through hypervisor control
  (Razzer, Snowboard)~\cite{razzer,snowboard}, instrumented delays (KRACE)~\cite{krace}, or
  \texttt{sched\_ext} scheduling (SECT)~\cite{sect2026}.
Exploits widen the window instead: Cai et al.\ slow path lookup and pause the victim, though the attack must still
  win the race~\cite{cai2009}, and DirtyCred and ExpRace (\chpt{6}) stall the kernel~\cite{lin2022dirtycred,exprace2021}.
rr replays a recorded run~\cite{rr2017}, and mutable replay carries it to changed code~\cite{viennot2013dora}; to
  reproduce a race, both need a recording of a run that hit it.

\noindent\textbf{Bisection.}
Flake-Aware Culprit Finding~\cite{Henderson2023FlakeAware} and Git Bayesect~\cite{GitBayesect} use Bayesian inference
  to locate a transition from noisy test outcomes.
Proposition~\ref{prop:bisect-bound} bounds the cost of that search under our hit/miss model.
Syzbot~\cite{syzkaller2026bisect} and Meerkat~\cite{meerkat2026} use repeated execution with flakiness safeguards;
  Meerkat also mutates PoCs to improve reproduction. These methods can classify a commit as unaffected after
  bounded non-reproduction.
SymBisect~\cite{symbisect2024} and LLMBisect~\cite{llmbisect2026} locate bug-introducing commits through
  symbolic analysis and patch/history reasoning, respectively.

\noindent\textbf{Version and patch analysis.}
VulScope~\cite{vulscope2021} migrates a crashing proof of concept across the versions of a program, and found 330
  versions that CVE and NVD entries did not list as vulnerable.
Chronos targets races, where a missed run decides nothing, and labels a version fixed only on positive evidence.
FIBER~\cite{fiber2018} and PDiff~\cite{pdiff2020} test whether a given patch is present in a binary, from the patch's
  code signature or a summary of what it changes.
Chronos runs the attack against the build.
A build that carries a patch can still be labeled vulnerable, as Debian~12's rpm package was, and a build fixed by
  different code can be labeled fixed, as \texttt{crun}'s and \texttt{youki}'s releases were.
Studies of development histories classify fixes that needed a later fix: supplementary bug
  fixes~\cite{park2012supplementary}, and incomplete or regressive security patches~\cite{li2017patches}.
Among concurrency bugs, some patches only made the bug rarer~\cite{lu2008mistakes}, which a natural run cannot tell
  from a fix.
Chronos labels each build in such a history, and each distribution package, by running the attack, so its labels
  show where a fix's record and a build disagree.

\newcommand{\rwY}{\tikz[baseline=-0.4ex]{\fill (0,0) circle[radius=0.45ex];}}
\newcommand{\rwP}{\tikz[baseline=-0.4ex]{\draw (0,0) circle[radius=0.45ex];\begin{scope}\clip (0,0)
  circle[radius=0.45ex];\fill (-0.45ex,-0.45ex) rectangle (0,0.45ex);\end{scope}}}
\newcommand{\rwN}{\tikz[baseline=-0.4ex]{\draw (0,0) circle[radius=0.45ex];}}
\newcommand{\rwNA}{--}
\newcommand{\rwH}[1]{\rotatebox{90}{#1}}
\newcommand{\rwG}[1]{\multicolumn{8}{@{}l}{\textit{#1}}\\}
\begin{table}[t]
  \caption{\textbf{Capabilities demonstrated by related systems and Chronos.}
    Tools differ in their target races, testing objectives and evidence semantics.}
  \label{tab:related-positioning}
  \centering
  \scriptsize
  \setlength{\tabcolsep}{2.2pt}
  \begin{tabularx}{\columnwidth}{@{}>{\raggedright\arraybackslash}Xccccccc@{}}
    \toprule
    & \rwH{Finds new races} & \rwH{Stock kernel, VM} & \rwH{Unmodified binary} & \rwH{Enforces order} &
      \rwH{Labels by effect} & \rwH{Needs fix evidence} & \rwH{Searches versions} \\
    \midrule
    \rwG{Name races in userspace programs}
    \textbf{Chronos} (syscall holds) & \rwP & \rwY & \rwY & \rwY & \rwY & \rwY & \rwY \\
    STING~\cite{sting2012} & \rwY & \rwN & \rwY & \rwY & \rwY & \rwP & \rwN \\
    \rwG{Process and kernel race reproduction}
    RacePro~\cite{racepro2011} & \rwY & \rwN & \rwY & \rwY & \rwP & \rwN & \rwN \\
    RRF~\cite{rrf2016} & \rwP & \rwP & \rwN & \rwP & \rwP & \rwN & \rwN \\
    SimEvo~\cite{simevo2017} & \rwY & \rwP & \rwY & \rwY & \rwP & \rwN & \rwP \\
    SysPro~\cite{SysPro2026} & \rwP & \rwY & \rwP & \rwP & \rwP & \rwN & \rwN \\
    CARDSHARK~\cite{cardshark2024} & \rwN & \rwN & \rwN & \rwP & \rwP & \rwN & \rwN \\
    \rwG{Instrumented thread-schedule testing}
    CONCURRIT~\cite{concurrit2013} & \rwP & \rwY & \rwP & \rwY & \rwP & \rwN & \rwN \\
    CHESS~\cite{chess2008} & \rwY & \rwY & \rwP & \rwY & \rwP & \rwN & \rwN \\
    \rwG{Races inside the kernel}
    Razzer~\cite{razzer} & \rwY & \rwN & \rwY & \rwY & \rwP & \rwN & \rwN \\
    KRACE~\cite{krace} & \rwY & \rwN & \rwN & \rwP & \rwP & \rwN & \rwN \\
    \rwG{Version search}
    Git Bayesect~\cite{GitBayesect} & \rwNA & \rwNA & \rwNA & \rwNA & \rwP & \rwN & \rwY \\
    Flake-Aware Culprit Finding~\cite{Henderson2023FlakeAware} & \rwNA & \rwNA & \rwNA & \rwNA & \rwP & \rwN & \rwY \\
    Meerkat~\cite{meerkat2026} & \rwN & \rwN & \rwY & \rwNA & \rwP & \rwN & \rwY \\
    \bottomrule
  \end{tabularx}

  \vspace{3pt}
  {\scriptsize\raggedright \rwY~shown in the cited source, \rwP~partly, \rwN~not shown, \rwNA~not applicable.
    Criteria for \rwP, by column. Finds new races: races from reports, or new exposure of a known race (in releases its
    advisories miss, other runtimes' releases, fixes that a variant of its attack bypasses, or packages a tracker lists as
    fixed).\par
    Stock kernel, VM: a kernel recorder, a simulator or a sanitizer kernel. Unmodified binary: run-time
    instrumentation.\par
    Enforces order: timed or random delays. Labels by effect: a failure, crash or detector report
    standing for the effect. Needs fix evidence: a defended result per attack. An open fix-evidence mark: a miss or
    pass counts as fixed, or no fixed verdict exists.\par}
\end{table}

\section{Conclusion}
\label{sec:conclusion}

Chronos turns the ad hoc timing and reproduction machinery built around various race exploits into a reusable Race-as-Code workflow.
Its fixed label needs positive evidence, so a lost race is never taken for a fix, and one held trial per commit
  replaces natural retries whose cost Proposition~\ref{prop:bisect-bound} bounds.
With its seccomp hold, one oracle serves bisection and CI regression testing
  without schedule search,
  timing calibration,
  a modified VM
  or a custom kernel.
With one new hold engine each, it also labeled two userspace memory races and two kernel use-after-frees.
The twenty-three-case corpus has definitive labels at both endpoints for twenty-one cases and vulnerable-side
  evidence only for two kernel cases.
Its labels helped us uncover three issues: two fixes that missed a path and a distribution package that kept one of
  them although its tracker called it fixed.
They also caught \texttt{runc}'s known reintroduction of an older CVE.

\ifdefined\chronosarxiv
\section*{Acknowledgment}
We thank Asad ur Rehman for his help with the technical review, proofreading, editing and presentation of the paper.
\section*{AI Usage Disclosure}
Under the authors' supervision, generative AI tools assisted with drafting and editing this preprint and with technical
  review, refinement, implementation and testing of the framework.
The research vision, problem formulation and key contributions originated with the authors, who led the framework's
  design.
\fi

\bibliographystyle{IEEEtran}
\bibliography{references}

\appendix
\renewcommand{\thesubsectiondis}{\Alph{subsection}.}

\subsection{Amplified Baseline and Bisection Protocols}
\label{sec:appendix-baseline}

The published CVE-2021-30465 proof of concept coordinates victims and swap loops.
We ran it without a hold against source-built \texttt{runc} on both sides of \texttt{0ca91f44} on a 4-CPU,
  Linux~6.17 host.
A \emph{launch} starts $N$ victim \texttt{runc create/start} containers against four \renameex{} swappers on one
  shared volume; it escapes if any victim reaches host \texttt{/}.
At $R{=}100$ launches per $N\in\{1,4,8,19,32\}$ and up to 30\,s of swapping, the per-victim rate fell from
  8.00\% ($N{=}1$) to 0.44\% ($N{=}32$); at $N{=}19$, 15/100 launches escaped without added load and 16/100 with
  four stress workers.
On a Linux~7.0, 16-vCPU cloud VM, the release binary with a 12\,s swap cap escaped in 70/100 launches at $N{=}4$
  and 84/100 at $N{=}19$ (Table~\ref{tab:natural-rates}).

The amplified bisection campaign used the same 19-victim/four-swapper topology on a 16-vCPU VM with source-built
  binaries: 100 vulnerable and 50 fixed calibration launches, then five $M{=}5$
  fixed-retry and five Git Bayesect walks (Table~\ref{tab:amplified-baseline}).

For the one-victim walks, $M$ is the number of misses after which a commit hitting at the calibration's lower
  Wilson bound~\cite{wilson1927} would pass as fixed with probability at most 0.05 (case D splits it over its six
  steps).
We clamped $M$ to the protocol's range of 20 to 150, which raised case A's 17 to 20 and held case D's 233 and the
  capped \texttt{crun} walks at 150.
The uncapped \texttt{crun} walks used the full 10{,}921.
The amplified walks' $M{=}5$ was set in advance from a 70\% hit-rate floor.

\paragraph{Fixed-retry protocol}
The one-victim fixed-retry campaign of \S~\ref{sec:eval-bisect} ran ten bisections with
  \texttt{git bisect run} at each $M$, on Linux~7.0.
In 22 of its 50 searches it took a vulnerable commit for fixed after $M$ misses and named a commit 1--17
  steps too early; the fixed commit never hit in 755 trials. A second campaign, on RQ1's setup, reached 10/10 only at
  $M{=}20$, with 157/562 (28\%) vulnerable hits.

The fixed-retry campaign's endpoint calibration measured 50 hits in 200 runs of the vulnerable version (25\%), against the 19\% its
  bisections' trials hit.
Five more 200-trial blocks with the same setup and classifier gave 15--21\% (173/1{,}000 overall),
  so the natural hit rate varied between batches.

\paragraph{Git Bayesect protocol}
In the Git Bayesect campaign of Table~\ref{tab:natural-rates}, the reference implementation~\cite{GitBayesect},
  which learns both hit rates itself, localized the fix at \texttt{0ca91f44} in all ten bisections on the CVE-2021-30465 narrow range
  with 79--306 classified natural trials each, beside 1{,}066 starts that failed before classification; Chronos used
  six trials per bisection, four midpoints and two endpoint checks.
We drove the \texttt{Bisector} class of Git Bayesect v1.2 directly, with its Jeffreys priors, on Linux~7.0.
The campaign's endpoint calibration measured 18 hits in 200 runs (9\%), against the 20\% its bisections' trials hit.
None of their 545 trials at fixed commits hit.
Each bisection stopped when one change point's posterior probability reached 0.95.

\begin{table*}[t]
  \caption{\textbf{Under Proposition~\ref{prop:bisect-bound}'s hit/miss model and rate assumptions, a natural-run
    bisector needs on average at least 32 to 10{,}317 trials for 95\% accuracy.} Chronos needed 6 to 16 trials.}
  \label{tab:bisect-races}
  \centering
  \scriptsize
  \setlength{\tabcolsep}{3pt}
  \begin{tabular*}{\textwidth}{@{\extracolsep{\fill}}llrrrrrr@{}}
    \toprule
    \textbf{Search} & \textbf{Range} & \textbf{Commits} & \textbf{Natural hits} & \textbf{Rate (\%)} &
      \textbf{Bound (runs)} & \textbf{Chronos} & \textbf{Bound / Chronos} \\
    \midrule
    \texttt{crun}, case A's attack, one victim & 0.11 to 0.12 & 67 & 13/2{,}000 & 0.65 [0.38, 1.11] &
      1{,}584 [927, 2{,}710] & 16$^{a}$ & 99 \\
    \quad amplified$^{b}$, 16 vCPUs & 0.11 to 0.12 & 67 & 102/400 & 25.5 [21.5, 30.0] & 37 [30, 44]
      & 16$^{a}$ & 2.3 \\
    \quad amplified$^{b}$, 4 vCPUs & 0.11 to 0.12 & 67 & 29/100 & 29.0 [21.0, 38.5] & 32 [23, 45]
      & 16$^{a}$ & 2.0 \\
    \quad walks' calibration$^{c}$ & 0.11 to 0.12 & 67 & 2/2{,}000 & 0.10 [0.03, 0.36] &
      10{,}317 [2{,}832, 37{,}624] & 16$^{a}$ & 645 \\
    D: CVE-2025-31133, fix & v1.2.7 to v1.2.8 & 32 & 17/200 & 8.5 [5.4, 13.2] & 96 [61, 153] & 7 & 14 \\
    \quad walks' calibration$^{c}$ & v1.2.7 to v1.2.8 & 32 & 8/200 & 4.0 [2.0, 7.7] & 207 [106, 409] & 7
      & 30 \\
    \texttt{youki}, case A's attack & v0.5.6 to v0.5.7 & 11 & 33/200 & 16.5 [12.0, 22.3] & 32 [23, 45] & 6 & 5.4 \\
    A: CVE-2021-30465, fix, narrow & rc94 to rc95 & 19 & 182/894$^{d}$ & 20.4 [17.8, 23.1] & 32 [28, 37] & 6 & 5.4 \\
    A: CVE-2021-30465, fix, wide & rc90 to rc95 & 1{,}267 & 182/894$^{d}$ & 20.4 [17.8, 23.1] & 82 [71, 94] & 12 & 6.8 \\
    \bottomrule
  \end{tabular*}

  \vspace{3pt}
  {\scriptsize\raggedright Natural hits: wins of the exploit run without the hold at the search's vulnerable
    endpoint, counting every attempt.
    Rate: with its 95\% Wilson interval.
    Bound: Proposition~\ref{prop:bisect-bound}'s bound at the rate and at its interval's ends, rounded down; Bound /
    Chronos uses the unrounded bound. It assumes
    the transition equally likely at any commit and the measured endpoint rate as a ceiling at every vulnerable
    commit (\S~\ref{sec:bisect}).
    Chronos: trials in its search, midpoints plus two endpoint checks
    (Table~\ref{tab:bisect-summary} and the release sweep).\par
    \smallskip
    \chNote{a}{Two searches of 8 trials, as in Table~\ref{tab:amplified-baseline}.}
    \chNote{b}{Each run is one launch of 19 victims and four swappers, which hits when any victim escapes; every
      launch at 0.11 also had refused victims.}
    \chNote{c}{The rate of record for the natural walks of Table~\ref{tab:amplified-baseline}.}
    \chNote{d}{Hits among the Git Bayesect campaign's trials at vulnerable commits
      (Table~\ref{tab:natural-rates}).}\par}
\end{table*}

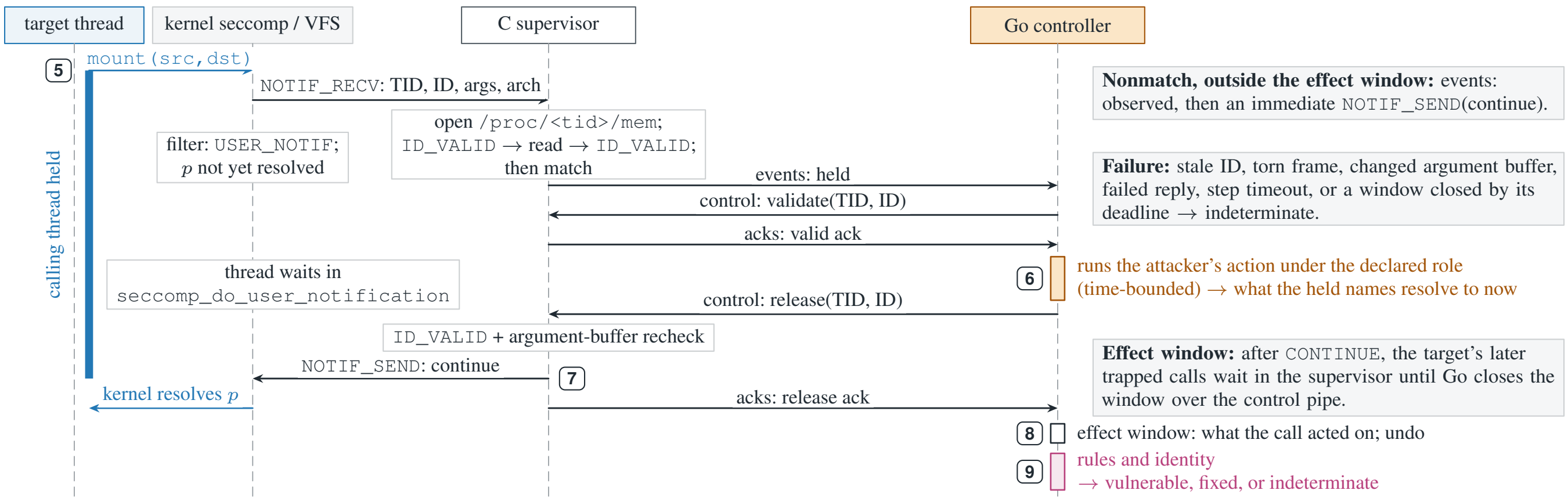
\begin{figure*}[t]
\centering
\resizebox{\textwidth}{!}{%
\begin{tikzpicture}[ch, x=1cm, y=1cm,
  every node/.style={font=\fontsize{6.5}{7.6}\selectfont},
  lanehead/.style={ch box, minimum height=0.42cm, inner ysep=1pt},
  msg/.style={fill=white, inner sep=1.2pt},
  proto/.style={->, line width=0.55pt, draw=ChInk},
  onlane/.style={ch box, align=center, inner xsep=3pt, inner ysep=1.5pt, fill=white},
  side/.style={ch box, fill=ChPaper, align=left, anchor=west, text width=5.2cm, inner xsep=3pt, inner ysep=2pt},
  act/.style={anchor=west, align=left, inner sep=0pt},
]
\def\xt{0.9}\def\xk{2.95}\def\xs{6.35}\def\xg{12.2}
\node[lanehead, draw=ChRuntime, fill=ChRuntimeTint, minimum width=1.6cm] at (\xt,5.5) {target thread};
\node[lanehead, fill=ChPaper, minimum width=2.3cm] at (\xk,5.5) {kernel seccomp / VFS};
\node[lanehead, draw=ChInkSoft, minimum width=2.0cm] at (\xs,5.5) {C supervisor};
\node[lanehead, draw=ChAttack, fill=ChAttackTint, minimum width=2.0cm] at (\xg,5.5) {Go controller};
\foreach \x in {\xt,\xk,\xs,\xg}
  \draw[ChGuide, densely dashed, line width=0.35pt] (\x,5.29) -- (\x,0.05);

\draw[ch held] (\xt+0.17,4.98) -- (\xt+0.17,1.44);
\node[text=ChRuntime, rotate=90, inner sep=0pt] at (\xt-0.22,3.21) {calling thread held};
\node[onlane] at (\xk,3.98) {filter: \texttt{USER\_NOTIF};\\$p$ not yet resolved};
\node[onlane] at (3.3,2.52) {thread waits in\\\texttt{seccomp\_do\_user\_notification}};

\draw[proto, draw=ChRuntime] (\xt+0.17,4.98) -- (\xk,4.98) node[midway, msg, above, text=ChRuntime] {\texttt{mount(src,dst)}};
\node[ch phase] at (\xt-0.18,4.98) {5};
\draw[proto] (\xk,4.64) -- (\xs,4.64) node[midway, msg, above] {\texttt{NOTIF\_RECV}: TID, ID, args, arch};
\node[onlane] at (\xs,4.13) {open \texttt{/proc/<tid>/mem};\\\texttt{ID\_VALID} $\to$ read $\to$ \texttt{ID\_VALID};\\then match};

\draw[proto] (\xs,3.66) -- (\xg,3.66) node[midway, msg, above] {events: held};
\draw[proto] (\xg,3.32) -- (\xs,3.32) node[midway, msg, above] {control: validate(TID, ID)};
\draw[proto] (\xs,2.98) -- (\xg,2.98) node[midway, msg, above] {acks: valid ack};

\fill[ChAttackTint] (\xg-0.08,2.84) rectangle (\xg+0.08,2.34);
\draw[ChAttack, line width=0.5pt] (\xg-0.08,2.84) rectangle (\xg+0.08,2.34);
\node[ch phase] at (\xg-0.32,2.59) {6};
\node[act, text=ChAttack] at (\xg+0.22,2.59) {runs the attacker's action under the declared role\\(time-bounded) $\to$ what the held names resolve to now};

\draw[proto] (\xg,2.18) -- (\xs,2.18) node[midway, msg, above] {control: release(TID, ID)};
\node[onlane] at (\xs,1.91) {\texttt{ID\_VALID} + argument-buffer recheck};
\draw[proto] (\xs,1.44) -- (\xk,1.44) node[midway, msg, above] {\texttt{NOTIF\_SEND}: continue};
\node[ch phase] at (\xs+0.27,1.44) {7};
\draw[proto] (\xs,1.1) -- (\xg,1.1) node[midway, msg, above] {acks: release ack};
\draw[proto, draw=ChRuntime] (\xk,1.1) -- (\xt+0.17,1.1) node[midway, msg, above, text=ChRuntime] {kernel resolves $p$};

\draw[ChInk, fill=white, line width=0.5pt] (\xg-0.08,0.92) rectangle (\xg+0.08,0.70);
\node[ch phase] at (\xg-0.32,0.81) {8};
\node[act] at (\xg+0.22,0.81) {effect window: what the call acted on; undo};
\fill[ChLabelTint] (\xg-0.08,0.58) rectangle (\xg+0.08,0.16);
\draw[ChLabel, line width=0.5pt] (\xg-0.08,0.58) rectangle (\xg+0.08,0.16);
\node[ch phase] at (\xg-0.32,0.37) {9};
\node[act, text=ChLabel] at (\xg+0.22,0.37)
  {rules and identity\\$\to$ vulnerable, fixed, or indeterminate};

\node[side] at (12.6,4.72)
  {\textbf{Nonmatch, outside the effect window:} events: observed, then an immediate \texttt{NOTIF\_SEND}(continue).};
\node[side] at (12.6,3.62)
  {\textbf{Failure:} stale ID, torn frame, changed argument buffer, failed reply, step timeout, or a window closed by
   its deadline $\to$ indeterminate.};
\node[side] at (12.6,1.45)
  {\textbf{Effect window:} after \texttt{CONTINUE}, the target's later trapped calls wait in the supervisor until Go
   closes the window over the control pipe.};
\end{tikzpicture}%
}
\caption{\textbf{A matched call is validated before the change, and its effect is observed before the target's next
  intercepted call proceeds.} Every failure branch makes the run indeterminate, and the thread (blue) stays in the
  kernel until \texttt{CONTINUE} (\chstep{7}). Boxed numbers are lifecycle phases (\S~\ref{sec:lifecycle}).}
\label{fig:chronos-freeze-protocol}
\end{figure*}

\begin{table*}[t]
  \caption{\textbf{All 23 injected faults of the supervisor's audit ended with the outcome declared before the run
    and left no supervisor or target process.}}
  \label{tab:notification-audit}
  \centering
  \scriptsize
  \setlength{\tabcolsep}{4pt}
  \begin{tabularx}{\textwidth}{@{}>{\raggedright\arraybackslash}p{0.15\textwidth}r>{\raggedright\arraybackslash}p{0.30\textwidth}Y@{}}
    \toprule
    \textbf{Group} & \textbf{Faults} & \textbf{Injected} & \textbf{Declared outcome} \\
    \midrule
    Repeated matching calls & 2 &
      Forty matching check calls, each released at once; four runnable peer threads during the capture &
      Each call is held once, in order, and the target completes; with the peers running, the capture is marked racy \\
    Cancellation and restart & 2 &
      Target killed while held; \texttt{SIGUSR2} with \texttt{SA\_RESTART} delivered while held &
      The killed target's notification fails validation: indeterminate (notification error); the restarted call is
      continued once and is not mistaken for the held call's return \\
    Path capture and argument integrity & 3 &
      Pathname split across two mappings; unterminated pathname ending at an unmapped page;
      protected pathname rewritten through \texttt{/proc/<tid>/mem} before release &
      The split path is captured exactly and released; the unterminated capture is dropped and the rewrite fails the
      re-read, both indeterminate \\
    Delay, transport, and protocol faults & 8 &
      Control delay within and beyond the controller's deadline, and above the supervisor's 5\,s limit; injected
      receive, event-publish, poll, and reply failures; a header plus one body byte, then a closed channel (torn frame) &
      The within-deadline exchange completes; the other seven are indeterminate, the torn frame at its body read \\
    Nonmatching calls and syscall mapping & 3 &
      Wrong pathname; a syscall outside the table; \texttt{stat} intercepted through \texttt{newfstatat} &
      The nonmatching calls run without a hold and the target completes; \texttt{newfstatat} is held and released
      before the use \\
    Controller labels and lifecycle failures & 5 &
      Vulnerable, fixed, and unattacked runs on the real supervisor; ordinary and sealed caller executables;
      effect-window deadline; observer delayed 100~ms; a failed supervisor stop, a supervisor failure after the window,
      and a failed target cleanup, each after a definitive observation &
      Vulnerable, fixed, and indeterminate as declared; the caller's executable is recorded; closing the effect window
      at its deadline leaves the effect unknown (indeterminate); the delayed observer runs once and decides; each later
      failure makes the run indeterminate \\
    \midrule
    \textbf{Total} & \textbf{23} & & \\
    \bottomrule
  \end{tabularx}
\end{table*}

\subsection{Statistical Bisection: Bound}
\label{sec:appendix-bayesect}

\begin{proof}[Proof sketch of Proposition~\ref{prop:bisect-bound}]
Fix the outcomes so far and the commit $c$ that the bisector tests next, and let $\Theta$ be the transition.
If $\Theta = \theta$, let the trial at $c$ hit with probability $p_{\theta,c}$: at most $q$, and zero when
  $c$ is fixed under $\theta$.
Rates may differ between commits.
Let $\mu$ be the mean of $p_{\Theta,c}$ under the current posterior.
Binary entropy is concave with $h(0) = 0$, so $h(p_{\theta,c}) \ge (p_{\theta,c}/q)\,h(q)$ for every $\theta$, and
  averaging over the posterior gives $\mathbb{E}[h(p_{\Theta,c})] \ge (\mu/q)\,h(q)$.
The trial's outcome $Y$ therefore carries
\[
  I(\Theta;Y) = h(\mu) - \mathbb{E}[h(p_{\Theta,c})] \le h(\mu) - \tfrac{\mu}{q}\,h(q) \le C(q)
\]
  bits, taking $x = \mu/q \in [0,1]$ in the definition of $C(q)$.
This one-step bound holds for every history of outcomes and every choice of the next commit.
Let $\tau$ be the bisector's stopping time, with $\mathbb{E}[\tau] < \infty$.
Pad the transcript with a fixed symbol after stopping.
The chain rule then bounds its information by
  $C(q)\sum_{t\ge1}\Pr(\tau\ge t)=C(q)\mathbb{E}[\tau]$~\cite{burnashev1976data}.
By Fano's inequality~\cite[\S2.10]{cover2006elements}, naming one of the $n$ commits chosen uniformly at random with
  error probability $\epsilon$ requires the outcomes to carry at least $\log_2 n - h(\epsilon) - \epsilon\log_2(n-1)$
  bits.
This adapts the classic lower bounds for noisy binary search~\cite{burnashev1974interval,benor2008bayesian}
  to noise that only turns hits into misses.
\end{proof}

\subsection{Supervisor Protocol and Fault-Injection Audit}
\label{sec:appendix-supervisor}

The seccomp-specific Fig.~\ref{fig:chronos-freeze-protocol} follows one held call through the supervisor and the controller, with its
  steps numbered by the lifecycle phases of \S~\ref{sec:lifecycle}.
The supervisor brackets its read of the held arguments with two \texttt{ID\_VALID} checks; after the change and
  release, it holds later intercepted calls until the effect is known.
It recognizes a restart by the same thread, arguments and instruction pointer, and observes return through the next
  trapped call, target exit or \texttt{/proc/<tid>/syscall}; undo follows observation.
An effect-window deadline makes the run indeterminate. A multithreaded capture is marked racy, but the mark alone does
  not change its label (\S~\ref{sec:timing-complete}).
All 23 injected protocol faults produced their declared outcome and no wrong definitive label
  (Table~\ref{tab:notification-audit}).

\subsection{Oracle-Check Sensitivity Audit}
\label{sec:appendix-oracle-ablation}

Table~\ref{tab:oracle-ablation} relabels 322 stored records of cases A--D with one oracle check removed
  at a time, without rerunning any target or attacker action.
The strict set has a held call and a recorded changed action: A 80, B 1 and D 60, with 141 definitive labels;
  C has no such run.
Each check that did not reject any run in the strict set also ran on 20 constructed records through the shipped decision
  procedure, with and without it.
The counterexample for positive fixed evidence is a vulnerable case-A mutant that mounts through an unheld
  destination.

\begin{table}[t]
  \caption{\textbf{Only removing object identity or positive fixed evidence changes recorded labels.}
    The other three checks did not reject any run in the strict set.}
  \label{tab:oracle-ablation}
  \centering
  \scriptsize
  \setlength{\tabcolsep}{4pt}
  \begin{tabularx}{\columnwidth}{@{}p{0.31\columnwidth}X@{}}
    \toprule
    \textbf{Check removed} & \textbf{Relabel outcome} \\
    \midrule
    Positive fixed evidence & Strict: 0 wrong; outside strict: 20 wrong fixed on the case-A mutant. \\
    Only the attacker changed the names & None failed; constructed: 20 of 20 definitive and wrong without it. \\
    Helper principal & None failed; constructed: 10 of 20 wrong without it, the other 10 indeterminate by
      identity. \\
    Object identity & Strict: 101 labels lost, 0 wrong; 40 rule-backed labels remain. \\
    Indeterminate on timeout & None timed out; constructed: 0 of 20 wrong without it, since identity and the
      held-change check catch them. \\
    \bottomrule
  \end{tabularx}
\end{table}

\begin{table*}[t]
  \caption{\textbf{Recorded evidence and the case premises needed to interpret it.}
    All definitive labels remain subject to the applicable failure, consistency and build checks.}
  \label{tab:evidence-contracts}
  \centering
  \scriptsize
  \setlength{\tabcolsep}{4pt}
  \begin{tabularx}{\textwidth}{@{}>{\raggedright\arraybackslash}p{0.12\textwidth}>{\raggedright\arraybackslash}p{0.14\textwidth}
    >{\hsize=1.05\hsize}Y>{\hsize=0.95\hsize}Y@{}}
    \toprule
    \textbf{Evidence} & \textbf{Cases} & \textbf{Recorded predicate and observation time} &
      \textbf{Case premise or boundary} \\
    \midrule
    Preserved: identity & Every row of Table~\ref{tab:cases} labeled preserved, or redirected when vulnerable &
      Candidate identities around the change; the family's observation of what the selected operation used after
      continuation. &
      The observer represents the declared use. The held object stands for the checked object unless the check
      was separately observed. \\
    Refusal & C and CVE-2019-19921: fixed side &
      No held use, abnormal non-timeout termination, and a declared diagnostic substring in terminal output.
      At an earlier syscall checkpoint, also a change completed while held and not seen reversed. &
      The diagnostic comes from the relevant check before prohibited effect. \\
    Call never made: omitted use & B, Cyrus and apport: fixed side &
      Normal non-timeout termination, no matched use, coverage and ambiguity checks, and no process of the target
      still running when its first one exits.
      At an instruction checkpoint, also the mechanism's record of the declared change, continuation and completion;
      at an earlier syscall checkpoint, a change completed while held. Neither may show the change reversed. &
      Covered operations and patterns include equivalent forbidden uses in this setup. \\
    Call never made: scoped absent read & CVE-2022-49919 and the iommufd use-after-free: fixed side &
      Source/function and selected post-checkpoint code inspection, with image and source identities. &
      The inspected scope covers the selected function and read pattern. \\
    Rule: final state & C, D and apport: vulnerable side &
      The declared property, such as namespace mount state, read before cleanup. &
      Excluding an earlier prohibited effect requires it to remain visible until observation. \\
    Rule: sanitizer & Apache, QEMU and the kernel cases: vulnerable side &
      Engine-specific matching of a report to the held function and available task, process or object fields
      after continuation. &
      The report is attributable to the selected use. Fixed requires a separate declared predicate. \\
    Rule: checked snapshot & Apache and QEMU: fixed side &
      Changed shared original, unchanged held copy, recorded return and selected-request output evidence. &
      The selected operand is the snapshot relevant to the declared use. \\
    Rule: confinement & Docker: fixed side &
      The selected task's root is measured at the hold. After target completion, it is compared with the declared
      area; constrained re-resolution is compared with candidate B, distinct from A.
      A satisfied confinement rule explains redirected identity only with normal termination. &
      Interpretation depends on observed use evidence or the case's binding between the candidate, selected use
      and retained state. The rule can decide when identity is silent. \\
    \bottomrule
  \end{tabularx}
\end{table*}

\subsection{Case Studies}
\label{sec:eval-cases}

\noindent\textbf{B: CVE-2025-23359 (NVIDIA hook).}\label{sec:eval-23359}
CVE-2025-23359~\cite{nvidiaCVE202523359} affects \texttt{libnvidia-container} through release~1.17.3.
While the hook's bind mount was held, one \renameex{} after the check caused the 1.17.3 hook to bind-mount a canary file at
  the exact compat destination, in 10 of 10 runs.
Without the swap, it did not.
Release 1.17.4 omits the compat mount, matching NVIDIA's fix, which stops mounting the container's compat libraries by
  default~\cite{nvidiaCVE202523359}.
Chronos bisected the toolkit's history from 1.17.3 to 1.17.4 to the commit ``Disable mounting of compat libs from
  container'' (Table~\ref{tab:bisect-summary}).
Two more searches located the commits that added an option to skip that mount and, later, an option to restore it;
  with that option, v1.18.0 was labeled vulnerable.

\noindent\textbf{C: CVE-2023-27561 (runc).}\label{sec:eval-27561}
Chronos's introduction search confirmed the boundary at the CVE-2021-30465 repair \texttt{0ca91f44}
  (Table~\ref{tab:bisect-summary}), which runc's fix \texttt{0abab45c} already names as the
  cause~\cite{runcFix202327561}; that repair reintroduced CVE-2019-19921, fixed in rc10~\cite{runcCVE201919921}
  (\S~\ref{sec:eval-revealed}).
The regression reached Ubuntu 20.04~\cite{ubuntuCVE202327561}: its \texttt{runc} build carries the backported
  repair, and the sweep labels that build vulnerable.

\noindent\textbf{D: CVE-2025-31133 (runc).}\label{sec:eval-31133}
Bisection found the fix \texttt{11cf22d1} (Table~\ref{tab:bisect-summary}), which validates \texttt{/dev/null} and mounts the masks through its opened
  descriptor.

\noindent\textbf{Two more container races.}
Chronos also labels the vulnerable and fixed releases of kubelet's subPath bind (CVE-2021-25741~\cite{nvdCVE202125741},
  v1.21.4 and v1.21.5) and of \texttt{runc}'s checks of its \texttt{/proc} mounts
  (CVE-2019-19921~\cite{runcCVE201919921}, rc9 and rc10).
Each of the 12 runs at these releases, three per endpoint, was labeled vulnerable or fixed; Table~\ref{tab:cases} counts each fix's exact parent and the fix.
Without the hold, their published exploits won 25 and 15 of 200 natural runs, respectively, on the vulnerable releases and none of
  50 on the fixed ones.

\noindent\textbf{Cyrus IMAP and Docker.}\label{sec:eval-cyrus-docker}
Cyrus IMAP's fixed label (SUSE bug 1241543) rests on a call never made: normal completion with no matching later
  \texttt{chmod} within the checked coverage, after the change completes while the \texttt{umask} checkpoint is held.

\looseness=-1
For Docker's CVE-2018-15664, held at the selected \texttt{open}, 18.09.6 was labeled vulnerable (redirected) in 3 of 3
  trials and 18.09.7 fixed under the declared confinement rule in 3 of 3.
Without the hold, the attack succeeded in 44 of 200 runs on 18.09.6 and none of 50 on 18.09.7.
The held and changed objects are what the path resolves to at the hold and after the change.
Its confinement rule (Table~\ref{tab:evidence-contracts}) compares the root measured at the selected task's hold
  with the declared area, then compares constrained re-resolution of the recorded path with the changed object,
  which differs from the held one.
A separate observation records what the selected use acted on.

\subsection{Evidence Predicates and Case Premises}
\label{sec:appendix-evidence-contracts}

Table~\ref{tab:evidence-contracts} separates what the implementation records from what the case must justify.
Rules report predicates to the shared decision procedure, with case premises determining their interpretation.

\subsection{Proof of Proposition~\ref{prop:placement}}
\label{sec:appendix-placement}

\begin{proof}
With the build, the initial state, the inputs and the target's other choices fixed, each step is a deterministic
  function of the complete state, including any time or memory the target reads; Chronos's held-name reads do not
  change it.
The premise lets the change move earlier past each step between the check and $U$, including other processes' steps,
  while the run stays legal and reaches the same complete state~\cite{lipton1975reduction,flanagan2005dpor}.
From any legal placement between the check and the use, move the whole change left to immediately after the check.
Each move preserves the complete state.
So~these~placements, including a held checkpoint within that interval, reach the use in the same state.
Thus $U$ acts on the same object, and a fixed label from preservation is correct at every placement.
A system-call footprint that avoids the changed names, objects and directories gives the premise only when no other
  dependency, such as a read of the time, separates the two orders.
\end{proof}

\subsection{Proof of Proposition~\ref{prop:bracket}}
\label{sec:appendix-bracket}

\begin{proof}
A linear fix search tests only commits between its current endpoints.
A definitive label replaces the endpoint that shares it, and an indeterminate run replaces neither, so by soundness
  the current vulnerable endpoint is a vulnerable revision and the current fixed endpoint a fixed one.
Along the path between them the value starts vulnerable and ends fixed, so it changes from vulnerable to fixed between
  some adjacent pair of commits.
With one transition in the range, revisions before it are vulnerable and those after it fixed, so it lies between
  the endpoints; when they are adjacent, they are that pair.
\end{proof}

\ifdefined\chronosfull\input{sections/supplement-head}\input{sections/appendix-extended}\clearpage\fi

\subsection{Ethics Considerations}
\label{sec:appendix-ethics}
We exercised vulnerabilities only in controlled local, cloud and CI environments, using experiment-created canaries
  and benign markers rather than third-party data or production targets.
Chronos's hold requires control of the target launch, which the attacker in our threat model does not have against a
  third party.
Before submission, we reported the chrony and Debian/rpm findings through their security channels.
Chrony's maintainers made the report public and posted a proposed fix~\cite{chronySocketReport2026,chronySocketFix2026};
  Debian amended its public tracker to identify the incomplete fix~\cite{debianRpmTracker2026}.

\end{document}